\documentclass[12pt,draftclsnofoot,journal,onecolumn]{IEEEtran}
\IEEEoverridecommandlockouts
\usepackage{mathtools}
\usepackage{enumitem} 
\usepackage{ifpdf}
\usepackage{amsthm}
\usepackage{subcaption}
\usepackage{amsmath}
\usepackage{nicefrac}
\usepackage{cite}
\usepackage{amsfonts}
\usepackage{comment}
 \usepackage{url}
 \usepackage{amssymb}
 \usepackage[font=small,labelfont=bf]{caption}
 \usepackage[paper=letterpaper,margin=0.75in]{geometry}
 \usepackage[ansinew]{inputenc}

\usepackage{lipsum}
\usepackage{hyperref}

\newtheorem{manualtheoreminner}{Theorem}
\newenvironment{mthm}[1]{%
	\renewcommand\themanualtheoreminner{#1}%
	\manualtheoreminner
}{\endmanualtheoreminner}

\theoremstyle{definition}

\newtheorem{fact}{Fact}
\newcommand*{\QEDB}{\hfill\ensuremath{\square}}%

\makeatletter
\renewenvironment{proof}[1][\proofname] {\par\pushQED{\qed}\normalfont\topsep6\p@\@plus6\p@\relax\trivlist\item[\hskip\labelsep\bfseries#1\@addpunct{.}]\ignorespaces}{\popQED\endtrivlist\@endpefalse}
\makeatother

\begin{document}
\title{Fundamental Limits of Invisible Flow Fingerprinting}

\author{
	\IEEEauthorblockN{Ramin Soltani\IEEEauthorrefmark{1},
		Dennis Goeckel\IEEEauthorrefmark{1}, Don Towsley\IEEEauthorrefmark{2}, and Amir Houmansadr\IEEEauthorrefmark{2}}
	
	\IEEEauthorblockA{\IEEEauthorrefmark{1}Electrical~and~Computer~Engineering~Department,~University~of~Massachusetts,~Amherst,
		\{soltani, goeckel\}@ecs.umass.edu\\}
	\IEEEauthorblockA{\IEEEauthorrefmark{2}College of Information and Computer Sciences, University of Massachusetts, Amherst,
		\{towsley, amir\}@cs.umass.edu}
	    
                       \thanks{ This work has been supported by the National Science Foundation under grants
                       CNS-1564067 and  CNS-1525642. The preliminary version of this work has been presented at the 51st Annual Asilomar Conference on Signals, Systems, and Computers,  November 2017~\cite{soltani2017towards}.

                   }
                        \thanks{This work has been submitted to the IEEE for possible publication. Copyright may be transferred without notice, after which this version may no longer be accessible.}
}

\date{}
\maketitle
\thispagestyle{plain}
\pagestyle{plain}

\newtheorem{definition}{Definition}

\begin{abstract}
Network flow fingerprinting can be used to de-anonymize communications on anonymity systems such as Tor by linking the ingress and egress segments of anonymized connections. Assume Alice and Bob have access to the input and the output links of an anonymous network, respectively, and they wish to collaboratively reveal the connections between the input and the output links without being detected by Willie who protects the network. Alice generates a codebook of fingerprints, where each fingerprint corresponds to a unique sequence of inter-packet delays and shares it only with Bob. For each input flow, she selects a fingerprint from the codebook and embeds it in the flow, i.e., changes the packet timings of the flow to follow the packet timings suggested by the fingerprint, and Bob extracts the fingerprints from the output flows. We model the network as parallel $M/M/1$ queues where each queue is shared by a flow from Alice to Bob and other flows independent of the flow from Alice to Bob. The timings of the flows are governed by independent Poisson point processes.
Assuming all input flows have equal rates and that Bob observes only flows with fingerprints, we first present two scenarios: 1) Alice fingerprints all the flows; 2) Alice fingerprints a subset of the flows, unknown to Willie. Then, we extend the construction and analysis to the case where flow rates are arbitrary as well as the case where not all the flows that Bob observes have a fingerprint. For each scenario, we derive the number of flows that Alice can fingerprint and Bob can trace by fingerprinting.
\end{abstract}

\section{Introduction}
\IEEEPARstart{G}{iven} the presence of communication systems in daily life and their rapid growth, e.g., cellular networks, internet of things, etc., security and privacy has emerged as a vital area of research and development~\cite{lopez2008wireless,nazanin_ISIT2017,nazanin_ISIT2018,hadian2016privacy,takbiri2018asymptotic,hadian2018privacy,nichols2001wireless,naghizadeh2016structural}.
 For every communication system, security involves not only allowing authorized users to communicate a message in a way that the message content is protected from unauthorized users, but also preventing access by malicious users. Hence, breaking the anonymity of users in an anonymous network such as Tor, Bitblinder, and Darknet plays a major role in preventing malicious use of technology. 

Even if the messages are encrypted, traffic analysis can be used to infer sensitive information from the packet characteristics such as timing patterns, sizes, and packet rates. For instance, packet timings can reveal information about passwords sent over SSH channels~\cite{song2001timing}. Also, traffic analysis can discover stepping stone attacks where malicious users employ compromised computers to relay their traffic~\cite{staniford1995holding,zhang2000detecting}. Furthermore, it can be used to find correlations between input and output links of a network to reveal connections between the links~\cite{syverson2001towards}.

Unlike passive traffic analysis which involves only recording traffic characteristics, such as packet timings, \emph{active} traffic analysis involves both recording and modifying traffic characteristics to embed information in them. For instance, in flow watermarking~\cite{houmansadr2009rainbow,houmansadr2011swirl,houmansadr2009multi}, watermarks are embedded into flows by changing their packet timings according to a unique secret pattern. Therefore, each flow contains one bit of information indicating whether it contains the watermark. However, in flow fingerprinting, the embedded patters are used to communicate information such as the identity of the party that performed fingerprinting~\cite{houmansadr2012design}, the location of the flow in the network where it was fingerprinted~\cite{soltani2017towards}, and the time when the fingerprint was embedded. Thus necessarily this will convey more than one bit of information. 

Active traffic analysis has emerged as a vibrant area of research recently. In~\cite{wang2003robust}, the authors propose detecting stepping stones using flow watermarking. Peng et al.~\cite{peng2006secrecy} show that this method is detectable and propose attacks on it. Wang et al.~\cite{wang2005tracking} show that the anonymity of VoIP calls made over an anonymity network can be broken using watermarking methods. Kiyavash et al.~\cite{kiyavash2008multi} propose a multi-flow attack on interval-based watermarking methods, which delay packets of specific intervals based on the value of the watermarks. Houmansadr et al. propose RAINBOW watermarking~\cite{houmansadr2009rainbow} and SWIRL~\cite{houmansadr2011swirl} which is a scalable traffic analysis method resilient against aggregated-flows attacks. They also study the capacity of flow watermarking~\cite{houmansadr2009channel} and propose a flow fingerprinting scheme allowing fingerprinting of millions of flows by perturbing the packet timings of relatively short lengths of flows~\cite{houmansadr2013need}. Rezaei et al.~\cite{rezaei2017tagit} introduce an active fingerprinting method called TagIt that works by slightly delaying packets into secret time intervals. In~\cite{wang2007network,yu2007dsss}, the authors consider watermarking and analyze invisibility and error probability of watermarking schemes in practice.

\begin{figure} 
	\centering
	\begin{subfigure}[b]{\textwidth}
		\includegraphics[width=\textwidth,height=\textheight,keepaspectratio]{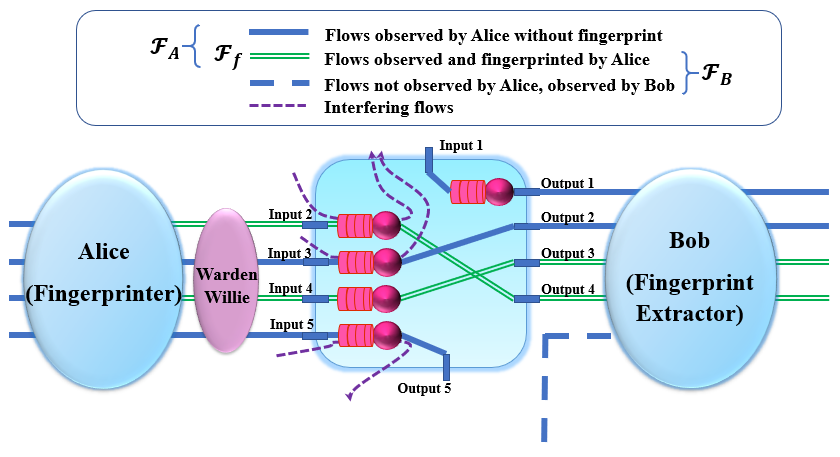}
		\caption{Setting 1: The network is modeled as independent parallel $M/M/1$ queues where each queue is shared between a flow from Alice to Bob (main flow) and other interfering flows that are independent of the main flow.}
		\label{fig:SysModa} 
	\end{subfigure}
	
	\begin{subfigure}[b]{\textwidth}
		\includegraphics[width=1\linewidth]{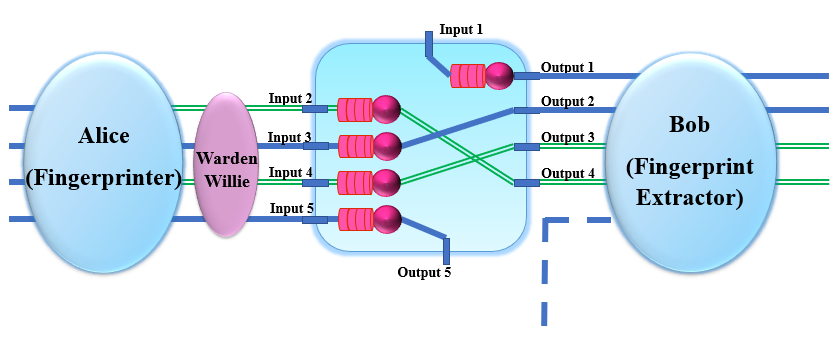}
		\caption{Setting 2: The network is modeled as independent parallel $M/M/1$ queues with single input/output where each queue conveys a flow from Alice to Bob.}
		\label{fig:SysModb}
	\end{subfigure}
	\caption[.]{Alice may fingerprint the flows, and Bob receives the fingerprinted flow after they pass through the network which adds timing noise to the fingerprints. Willie who is warden of the network protects the links from being traced; he wishes to determine whether Alice has fingerprinted flows.}\label{fig:SysMod}
\end{figure}

Previous active traffic analysis methods do not offer theoretical guarantees on the trade-off between performance (number of the flows) and~\textit{invisibility}, i.e., altering the packet timings so that the outcome is statistically indistinguishable from intact packet timings. When the traffic analyzer is the warden of the network who protects the links from being traced by anonymous users (e.g., for de-anonymization), invisibility of traffic analysis is important since attackers (anonymous users) can evade analysis if they are aware of the fingerprinting process. Even when the traffic analyzer is not the network warden, the invisibility of the traffic analysis is crucial in order to hide from the network warden. In this paper, we consider invisible fingerprinting to trace the input and output links of a network in the presence of a network warden. Consider an anonymous network where connections between input and output links are unknown. We model the network as $M$ parallel work conserving queues with Poisson arrivals and exponential service times ($M/M/1$ queues) and First In First Out (FIFO) discipline. Queues are independent and each queue is shared by a flow from the input of a network to the output of the network and other flows independent of the flow from Alice to Bob (see Fig.~\ref{fig:SysModa}). Alice has access to the input flows and she can buffer and release packets when she desires. On the other side of the network, Bob has access to the output flows so he can read the packet timings of the flows. Alice and Bob wish to perform fingerprinting to infer the connections between input links and output links, without being detected by Willie whose goal is to discover flow fingerprints. 

We consider the following problem: in a time interval of length $T$, can Alice and Bob perform fingerprinting to link input and output flows of the network without being detected by Willie, and if yes, how can they do so and what is the maximum number $m$ of flows that they can link reliably? For the case where packet timings of each flow is an independent instantiation of a Poisson process, we present the construction and analysis, and calculate the asymptotic expression for $m$ as a function of $T$. We first assume flow packet rates are equal and that Bob observes only flows with fingerprints and consider two main scenarios: 1) Alice fingerprints all flows she observes; 2) Alice fingerprints a subset of the flows, and the subset is unknown to Willie. Then, we present the extensions to arbitrary flow rates as well as the case where Bob observes a set of flows in which not all flows are fingerprinted.  

The contributions of this work relative to the conference version in~\cite{soltani2017towards} are:
\begin{itemize}
\item For the case where Alice fingerprints all flows, we present more details of the analysis for both the reliability and the number of possible fingerprinted flows.
\item For the case where Alice fingerprints a subset of the flows, in~\cite[Theorem 2]{soltani2017towards}, we presented a scenario where Alice fingerprints each flow independently with probability $q$. Here, we present a slightly different variation of this scheme where instead of a probabilistic selection of flows for fingerprinting, Alice fingerprints a subset of the flows which is known to both Alice and Bob (see Theorem~\ref{thm:pflows}). Furthermore, the results of~\cite[Theorem 2]{soltani2017towards} were applicable only for specific values of $q$ and the total number of flows that yield a close to a maximal number of traceable flows. Here, we present the results for arbitrary $q$ and number of flows (see Theorem~\ref{thm:pflowsw2}). 
\item The extension to the case of arbitrary flow rates was discussed in~\cite[Section V.B]{soltani2017towards} briefly. Here, we present the full construction and analysis  (see Theorems~\ref{thm:allflowsd} and~\ref{thm:pflowsd}).
\item We analyze the case where Bob observes a set of flows in which some of them are not fingerprinted. We present a construction where Bob uses a detector to determine if a flow is fingerprinted (see Theorems~\ref{thm:allflowsw} and~\ref{thm:pflowsw}).
\item We present simulation results for Willie's probability of error, the probability that Alice runs out of packets, Bob's probability of error, and robustness of our scheme against changes in processing time of queues.  
\end{itemize}

The remainder of the paper is organized as follows. We present the system model, definitions, and invisibility and reliability metrics employed in this paper in Section~\ref{prerequisites}. Then, in Sections~\ref{scen1} and~\ref{scen2}, we present constructions and analyses for the two main fingerprinting scenarios. In Section~\ref{sec3}, we present the extensions of the main scenarios to arbitrary flow rates, and in Section~\ref{sec4}, we present the extensions of the main scenarios to the case where Bob observes flows with and without fingerprints. Section~\ref{disc} discusses the results, and Section~\ref{sec:fw} discusses future work. We conclude in Section~\ref{sec:con}.
\section{System Model, Definitions, and Metrics}~\label{prerequisites}

\subsection{System Model} \label{sec:sysmod}
We consider a set of $M$ flows between $M$ pairs of input and output links. We assume the links are known but not the pairings. Also present are two parties Alice and Bob whose goal is to identify some or all of the pairings by fingerprinting, without a third party, Willie, detecting this identification. Moreover, Alice and Bob wish to do so within the time interval $[0,T]$. Alice, Bob, and Willie know that all packet timings are governed by Poisson processes and they the rate of each flow that they observe. 

Alice has access to a subset of the input links where each link conveys a packet flow $f_i^{(A)}\in\mathcal{F}_{A}=\{ f_1^{(A)},f_2^{(A)},\ldots,f_{M}^{(A)}\}$. She is allowed to buffer packets and release them from her buffer but no other operations (e.g., inserting packets, changing packet ordering). Willie is located between Alice and the network, and he watchfully observes all of the input links accessed by Alice ($\mathcal{F}_{A}$) to detect whether or not Alice is fingerprinting flows (see Fig.~\ref{fig:SysMod}). Willie is able to verify the sources and the order of the packets. Therefore, if Alice inserts a packet of her own or re-orders the packets on any of the links to transmit information to Bob, Willie will detect her immediately. Bob observes a subset of the output links where each link conveys a packet flow $f_j^{(B)}\in\mathcal{F}_{B}= \{f_1^{(B)},f_2^{(B)},\ldots,f_{M_{b}}^{(B)}\}$. He is only allowed to observe the time of the arrival of each of the packets in each flow. Bob and Willie cannot manipulate the flows (e.g., change packet timings, remove packets, insert packets, change packet ordering). 
 
Prior to fingerprinting, Alice generates a codebook of fingerprints and shares it with Bob. The codebook is secret, and thus Willie does not have access to it. On the other side of the network, Bob uses the codebook to extract the fingerprints and identify the flows. 

Each fingerprint (codeword) of the codebook corresponds to a sequence of inter-packet delays, which plays the role of a unique flow identifier. Alice embeds a unique fingerprint in each flow, i.e., she buffers packets of each flow and releases them according to timings associated with a fingerprint. We denote by $\mathcal{F}_{f}\subset \mathcal{F}_{A}$ the set of flows with fingerprints. In general, not every fingerprinted flow is observed by Bob. However, since our goal is to calculate the maximum number of flows that can be traced by Alice and Bob, we assume Bob observes all fingerprinted flows, i.e., $\mathcal{F}_{f}\subset \mathcal{F}_{B}$.

As Willie is only able to read the channel, he cannot change packet timings; however, packet timings change after they pass through the network. Nevertheless, we present a construction where Bob can successfully identify the flows.   

We model the network as $M$ parallel First In First Out (FIFO) queues with exponential service times ($M/M/1$ queues). We consider two settings for the network:

\begin{enumerate}
	\item Setting 1: each $M/M/1$ queue is shared by the flow Alice and Bob are monitoring, which we refer to it as ``main flow'', and other flows independent of the main flow, which we refer to them as ``interfering flows''. (see Fig.~\ref{fig:SysModa}).
	\item Setting 2: each $M/M/1$ queue conveys just the flow Alice and Bob are monitoring (see Fig.~\ref{fig:SysModb}).
\end{enumerate}
Denote by $q_i$ the $i^{\mathrm{th}}$ queue, and by $\mu_i$, $\lambda_i$, and $\lambda'_i$ the service rate, the input rate, and the sum of the rates of the interfering flows at $q_i$, respectively. We term $\mu'_i=\mu_i-\lambda'_i$ the effective service rate~\cite{liutiming} of $q_i$ and we assume Alice knows the effective service time of all queues $q_1,\ldots,q_M$. The queues are stable, i.e., $\lambda_i + \lambda'_i< \mu_i$.

First, we consider Setting 1 (shown in Fig.~\ref{fig:SysModa}). Assuming the flow rates of the flows observed by Alice and Bob are the same ($\lambda_i=\lambda$) and that Bob observes only the set of fingerprinted flows ($\mathcal{F}_{B}=\mathcal{F}_f$), we present two scenarios:
\begin{itemize}
	\item Scenario~1 (analyzed in Section~\ref{scen1}): Alice fingerprints all flows to which she has access ($\mathcal{F}_{f}=\mathcal{F}_{A}$). 
	\item Scenario~2 (analyzed in Section~\ref{scen2}):	Alice fingerprints a subset of the flows to which she has access ($\mathcal{F}_{f}\subset\mathcal{F}_{A}$).
\end{itemize}
\noindent Then, considering the same setting for the network (Setting 1 shown in Fig.~\ref{fig:SysModa}), we present Scenarios~3 and~4 which are extensions of Scenarios~1 and~2, respectively, to the case that flow rates are arbitrary. Scenarios~3 and~4 are analyzed in Sections~\ref{scen3} and~\ref{scen4}, respectively. Next, we consider Setting 2 (shown in Fig.~\ref{fig:SysModb}) and present Scenarios~5 and~6, which are extensions of Scenarios~1 and~2, respectively, to the case that Bob observes fingerprinted flows as well as other flows that are not fingerprinted ($\mathcal{F}_f\subset \mathcal{F}_{B}$). If Bob observes a flow $f_i^{(B)}$ that is not fingerprinted, the flow can be either coming from Alice ($f_i^{(B)} \in \mathcal{F}_{A}$) or other inputs of the network ($f_i^{(B)} \notin \mathcal{F}_{A}$). Scenarios~5 and~6 are analyzed in Sections~\ref{scen5} and~\ref{scen6}, respectively. We show that in each scenario Alice can fingerprint the flows invisible to Willie but distinguishable by Bob. In addition, we determine the number of flows that Alice and Bob can invisibly and reliably trace by fingerprinting. 

Next, we present definitions and describe invisibility and reliability metrics. 
\subsection{Definitions}
Willie uses hypothesis testing to detect whether Alice is fingerprinting:
\begin{itemize}
	\item $H_0$: Alice is not fingerprinting. 
	\item $H_1$: Alice is fingerprinting.
\end{itemize}
Denote $\mathbb{P}_{\mathrm {FA}}$ as the false alarm probability of rejecting $H_0$ when Alice is not fingerprinting (type I error), and $\mathbb{P}_{\mathrm {MD}}$ as the missed detection probability of rejecting $H_1$ when Alice is fingerprinting (type II error). To give more power to Willie, we assume he knows the probability that Alice is fingerprinting, $\mathbb{P}(H_1)=1-\mathbb{P}(H_0)$.

Similar to the definition of covertness~\cite{soltani2014covert,soltani2015covert,soltani2016allerton,soltani2018covert,soltani2018allerton}, we define invisibility\cite{soltani2017towards}:

\begin{definition}\label{def:inv} (Invisibility) Alice's fingerprinting is {\em invisible} (covert) if and only if she can lower bound Willie's probability of error, $\mathbb{P}_{\mathrm e}^{(\mathrm w)}= \frac{\mathbb{P}_{\mathrm {FA}}+\mathbb{P}_{\mathrm {MD}}}{2}$, by  $\frac{1}{2}-\epsilon$ for any $\epsilon>0$, as $T \to \infty$. We term $\epsilon$ the invisibility parameter. 
\end{definition}
\begin{definition} \label{def:rel} (Reliability) Alice's fingerprinting is {\em reliable} if and only if for any $\zeta>0$ and any flow, the probability of the failure event satisfies
$\mathbb{P}_{\mathrm{f}}\leq \zeta$ as $T \to \infty$. We term $\zeta$ the reliability parameter. For a flow with a fingerprint the failure event occurs when one of the following events occurs:
\begin{itemize}
\item Alice cannot successfully fingerprint the flow since she does not have a packet available to release when she needs one. We denote by $\mathbb{P}_{\mathrm{f}_1}$ the probability of this event.
\item Bob cannot extract the fingerprint successfully. We denote by $\mathbb{P}_{\mathrm{f}_2}$ the probability of this event.
\end{itemize}
For a flow without a fingerprint, the failure event occurs when Bob detects a fingerprint. We denote by $\mathbb{P}_{\mathrm{f}_3}$ the probability of this event.
\end{definition}
Note that both $\mathbb{P}_{\mathrm{f}_3}$ and $\mathbb{P}_{\mathrm{FA}}$ refer to the (erroneous) detection of fingerprints when flows are not fingerprinted; however, the former refers to detection by Willie after observing all the flows, and the latter refers to detection by Bob for each flow.

\begin{definition} \label{def:3} (Lambert-W function) The Lambert-W function is the inverse function of $f(W)=We^W$.
\end{definition}

We present results under the assumption that $\mathbb{P}(H_0)=\mathbb{P}(H_1)=1/2$. We show in Appendix~\ref{ap.1} that this results in invisibility for the general case where $\mathbb{P}(H_0)\neq\mathbb{P}(H_1)$. In this paper, we use standard Big-O, Little-o, Big-Omega, little-omega, and Big-Theta notations~\cite[Ch. 3]{cormen2009introduction}.

\section{Scenario~1: All flows are fingerprinted, Setting 1} \label{scen1}
Consider Scenario~1: Alice fingerprints all flows she observes ($\mathcal{F}_{f}=\mathcal{F}_{A}$), and Bob observes only the fingerprinted flows ($\mathcal{F}_{B}=\mathcal{F}_f$). All of flow rates are equal ($\lambda_i=\lambda$). We consider Setting 1 (see Fig.~\ref{fig:SysModa}), i.e., $M$ parallel $M/M/1$ queues where each queue is shared by a fingerprinted flow and other interfering flows independent of the fingerprinted flow. Alice fingerprints the input flows during time interval $[0,T]$, and Bob extracts the fingerprints from the flows on the output links of the network to infer the connections between input and output flows. 

\begin{figure}
	\begin{center}
		\includegraphics[keepaspectratio]{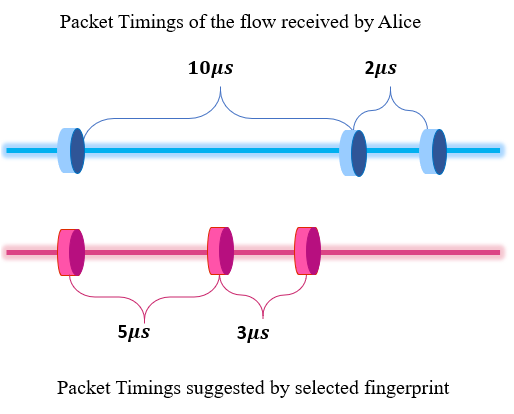}
	\end{center}
	\caption{An example of when Alice cannot successfully fingerprint a flow: the packet timings of the flow received by Alice and the packet timings suggested by the selected fingerprint are $[10\mu s,2\mu s\ldots]$ and $[5 \mu s,3 \mu s,\ldots]$, respectively. Alice faces a causality problem when she needs to send the second packet since she has to send it before she receives it.}
	\label{fig:timings}
\end{figure}

Alice buffers packets and releases them according to a fingerprint. She uses a secret codebook where each codeword (fingerprint) is a unique flow identifier consisting of a sequence of inter-packet delays. Because the timings of packets that Alice receives as well as the codewords are random, Alice will face a causality problem: the need to send a packet before she receives it. We give an example of when Alice cannot successfully fingerprint a flow in Fig.~\ref{fig:timings}. 

Consider a flow and assume the inter-arrival times of this flow before Alice makes any changes are $[10\mu s,2\mu s\ldots]$. Also assume Alice selects a fingerprint $C(W)=[5 \mu s,3 \mu s,\ldots]$ from her codebook. Note that the inter-arrival time between the first and second packets of the flow is $10\mu s$ but Alice has to alter the packet timings of the flow to achieve an inter-arrival of $5\mu s$ between the first and the second packets. In other words, she has to send the second packet before she receives it.

To account for this, prior to fingerprinting, Alice invisibly slows down the flow in order to buffer packets~\cite[Section IV]{soltani2015covert}. This ensures she will have a packet in her buffer to transmit at the appropriate times and can fingerprint the flow successfully.

We calculate the number of flows $m=M$ that Alice and Bob can trace by fingerprinting using this scheme, asymptotically as a function of $T$. 
\begin{mthm}{1} \label{thm:allflows} 
	Consider Setting~1 (see Fig.~\ref{fig:SysModa}). If Alice fingerprints all $M$ input flows ($\mathcal{F}_f=\mathcal{F}_{A}$) whose rates are equal ($\lambda$) and Bob only observes fingerprinted flows ($\mathcal{F}_{B}=\mathcal{F}_f$), then Alice and Bob can invisibly and reliably trace $m=M=\mathcal{O}(T/\log T)$ flows in a time interval of length $T$.
\end{mthm}
\begin{proof}
\textbf{Construction}: Per above, Alice uses a scheme consisting of two phases of lengths $T_1$ and $T_2$, and employs a codebook of fingerprints to embed in the flows. The codebook construction is similar to the one adopted in~\cite{soltani2017towards,soltani2015covert,soltani2016allerton}. In particular, Alice generates $m$ independent instantiations of a Poisson process with parameter $\lambda T_2$, where $T_2$ is the length of the second phase, as follows. To generate the $l^{\mathrm{th}}$ codeword ($1\leq l \leq m$), first a number $n_l$ is generated according to a Poisson distribution with mean $\lambda T_2$, and then $n_l$ points are distributed randomly and uniformly in a time interval of length $T_2$~\cite{verdubitsq} (see  Fig.~\ref{fig:codebook}). Therefore, the codebook contains $m$ fingerprints (codewords) $\{C(W_l)\}_{l=1}^{l=m}$. Alice selects a fingerprint for each flow and applies the inter-packet delays of the chosen fingerprint to the packets of the flow. The codebook is shared with Bob, not know to Willie.  

Alice divides the time interval of length $T$ into two phases (see Fig.~\ref{fig:Twophased}): 
\begin{itemize}
	\item Phase 1 (buffering phase) of length $T_1$: Alice slows each flow from rate $\lambda$ to rate $\lambda-\Delta$ to buffer packets, i.e., if she receives a packet at time $\tau$, she transmits it at time $ \frac{\tau \lambda}{\lambda-\Delta}$. This allows her to build up a backlog of packets in her buffer which ensures that she will be able to fingerprint each flow during the next phase successfully. 
	\item Phase 2 (fingerprinting phase) of length $T_2=T-T_1$: for each flow, she selects a fingerprint from her codebook and then alters the packet timings of the flow according to the selected fingerprint.
\end{itemize}
\begin{figure}
\begin{center}
\includegraphics[width=\textwidth,height=\textheight,keepaspectratio]{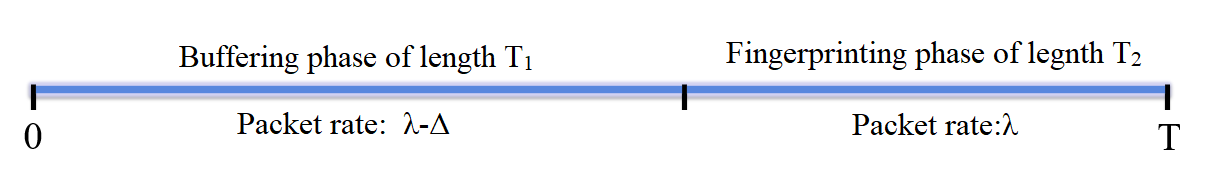}
\end{center}
 \caption{Alice's divides the time interval of length $T$ into two phases: a buffering phase of length $T_1$ where packets of each flow are slowed down, and a fingerprinting phase of length $T_2=T-T_1$ where Alice fingerprints the flows.}
 \label{fig:Twophased}
 \end{figure}
\noindent The lengths of the two phases are,
\begin{align}
\label{eq:lph10} T_1&= \frac{ T m\alpha}{1+m\alpha},\\
\label{eq:lph20}  T_2&=T-T_1= \frac{T}{1+m\alpha},
\end{align}
\noindent where $\alpha$ is a constant defined later, and $m$ is the number of flows to be fingerprinted.

\begin{figure}
\begin{center}
\includegraphics[width=\textwidth,height=\textheight,keepaspectratio]{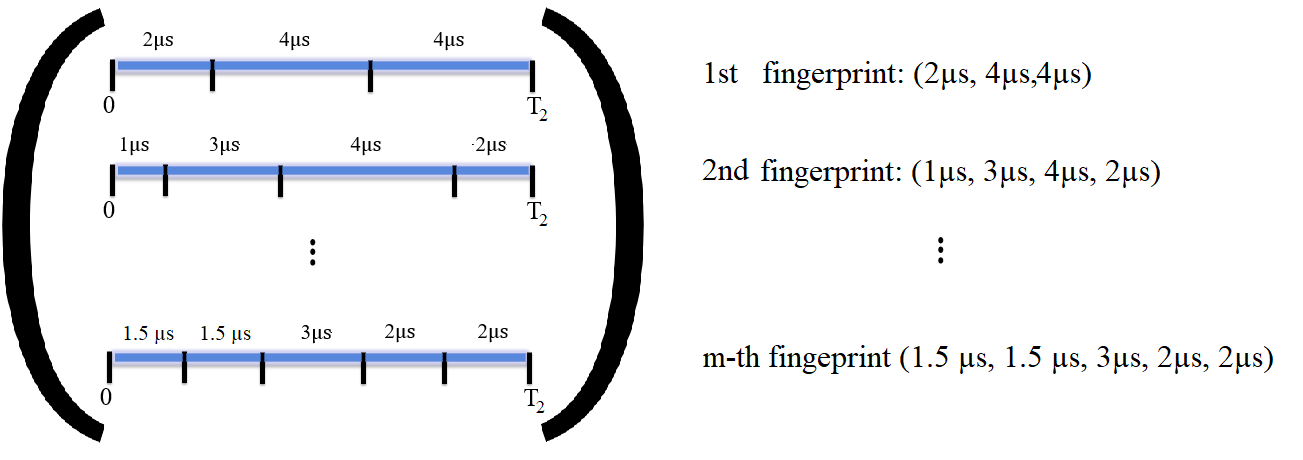}
\end{center}
 \caption{Codebook generation: Alice generates a codebook whose codewords (fingerprints) specify the sequence of inter-packet delays to be embedded in the flows. Each codeword is an instantiation of a Poisson process of rate $\lambda_{\min}=\min(\lambda_1,\ldots,\lambda_m)$ in a time interval of length $T_2$. For each codeword, first a random variable $N$ is generated according to the Poisson distribution with parameter $\lambda T_2$. Then $N$ points are placed uniformly and randomly in the time interval of length $T_2$. The codebook is shared with Bob, but it is unknown to Willie.}
 \label{fig:codebook}
 \end{figure}
 
\textbf{Analysis}: ({\em Invisibility}) Similar to the analysis of covertness in~\cite[Theorem 2]{soltani2015covert}, we can show that Alice's fingerprinting is invisible. Consider the first phase. We can show that for all $\epsilon\in (0,\frac{1}{2})$, Alice can slow down the flows from rate $\lambda$ to rate $\lambda-\epsilon \sqrt{{2 \lambda}/{m T_1}}$, and achieve (see the proof in Appendix~\ref{ap.2})
\begin{align}
\label{eq:3} \mathbb{P}_{\mathrm e}^{(\mathrm w)}>\frac{1}{2}-{\epsilon},
\end{align}
\noindent  where $\mathbb{P}_{\mathrm e}$ is Willie's error probability. Thus, her buffering is invisible. In the second phase, the packet timings for each flow is an instantiation of a Poisson process with rate $\lambda$ and hence the traffic pattern is indistinguishable from the pattern that Willie expects to observe. Hence, the scheme is invisible.

({\em Reliability}) Now, we show that Alice's fingerprinting satisfies all of the conditions in Definition~\ref{def:rel}, and thus is reliable. Note that all flows have fingerprints. By the union bound:
\begin{align}
\label{eq:uni}\mathbb{P}_f \leq \mathbb{P}_{\mathrm{f}_1}+\mathbb{P}_{\mathrm{f}_2}.
\end{align}
Thus, to show the fingerprinting is reliable, it suffices to show that $\mathbb{P}_{\mathrm{f}_1}+\mathbb{P}_{\mathrm{f}_2}\leq \zeta$ for all $\zeta>0$.

First, we show that $\mathbb{P}_{\mathrm{f}_2}\to 0$ as $T \to \infty$ for each flow, i.e., Bob can successfully extract a fingerprint from each flow. Recall that Alice fingerprints all $m$ flows that she observes and Bob observes only the flows fingerprinted by Alice ($\mathcal{F}_{A}=\mathcal{F}_{f}=\mathcal{F}_{B}$). Therefore, $m=|\mathcal{F}_{A}|=|\mathcal{F}_{f}|=|\mathcal{F}_{B}|$, where $|\cdot|$ denotes the cardinality of a set. 

Without loss of generality, we assume that flow $f_i^{(B)}$ passes through the $i^{\mathrm{th}}$ queue ($q_i$). Denote by $C_i$ the capacity of $q_i$ for the transmission of information via packet timings. Recall that $q_i$ is an $M/M/1$ queue with multiple inputs and outputs and that Alice establishes a timing channel on each input flow to send a fingerprint to Bob. Recall that $q_i$ is an $M/M/1$ queue with multiple inputs and outputs and that Alice establishes a timing channel on each input flow to send a fingerprint to Bob. Therefore, we use the bound on the capacity of the timing channel for a shared $M/M/1$ queue~\cite[Proposition 1]{liutiming}:
\begin{align}
\label{eq:capq}C_i\geq \lambda \log{\left({(\mu_i-\lambda'_i)}/{\lambda}\right)},
\end{align}
\noindent where $\lambda'_i$ is the sum of rates of the interfering flows passing through $q_i$, and $\mu_i$ is the service rate of $q_i$. Note that~\eqref{eq:capq} implies that although $q_i$ changes the packet timings of the flow and thus the embedded fingerprint, Bob is able to successfully decode at least $C_i$ nats/second bits from the packet timings of the flow and thus extract Alice's fingerprint. From~\cite[Definition 1]{verdubitsq}, the rate of the codebook is  $\frac{\log{m}}{T_2}$, and ~\cite[Definition 2]{verdubitsq},~\eqref{eq:capq} implies that all transmission rates smaller than $\lambda \log{\left({(\mu_i-\lambda'_i)}/{\lambda}\right)}$ result in a decoding error probability that tends to zero as $T_2 \to \infty$. Therefore, we require
\begin{align}
\label{eq:22} \frac{\log m}{T_2}< \lambda \log{\left({(\mu_i-\lambda'_i)}/{\lambda}\right)}
\end{align}
\noindent for Bob to successfully extract the fingerprint from $f_i^{(B)}$. Note that~\eqref{eq:22} holds for all $1\leq i\leq m$. Hence, as long as 
\begin{align}
\label{eq:cap} \frac{\log m}{T_2}< C,
\end{align}
\noindent where 
\begin{align}
\label{eq:c} C=\lambda \log{\left({\underset{i}{\mathrm{min}} \{\mu_i-\lambda'_i\}}/{\lambda}\right)},
\end{align}
\noindent for each flow $\mathbb{P}_{\mathrm{f}_2} \to 0 \text{ as } T_2 \to \infty$. Note that~\eqref{eq:lph20} implies that $T_1,T_2 \to \infty$ as $T \to \infty$. Therefore,
\begin{align}
\label{eq:112}\mathbb{P}_{\mathrm{f}_2} \to 0 \text{ as } T \to \infty.
\end{align}

Next, we show that $\mathbb{P}_{\mathrm{f}_1} \leq \zeta$, i.e., Alice can successfully fingerprint the flows. Recall that Alice accounts for the causality problem by buffering packets before she starts fingerprinting. Since in the first phase Alice slows down the packet rate from rate $\lambda$ to rate $\lambda-\epsilon \sqrt{{2 \lambda}/{m T_1}}$, on average she can buffer $\epsilon \sqrt{{2 \lambda T_1 }/{m}}$ packets. Consequently, we can apply the weak law of large numbers (WLLN) to show that the probability that Alice buffers more than $\epsilon \sqrt{{ \lambda T_1 }/{m}}$ packets tends to one, as $T$ tends to infinity. Now, we have to answer this question: noting that Alice has $\epsilon \sqrt{{ \lambda T_1 }/{m}}$ packets in her buffer, what is the probability that Alice cannot successfully fingerprint $f_i^{(A)}$?

Because Alice receives and transmits packets on each flow according to two independent Poisson processes of rate $\lambda$, and the Poisson process is memoryless, we model the process as a symmetric random walk on a 1-D grid to answer this question \cite{soltani2015covert}. 
The location of the walker corresponds to the number of packets in Alice's buffer. The walker goes from location $z$ to $z+1$ when Alice receives a packet, and goes from location $z$ to $z-1$ when Alice transmits a packet. Denote by $\mathbb{P}_{k,t}$ the probability of the event that the walker starting from the location $z=k$ reaches the point $z=0$, at least once, during the time $[0,t]$. Then~\cite[Eq. (27)]{soltani2015covert}:
\begin{align}
\label{eq:125}\lim\limits_{t \to \infty}\mathbb{P}_{k,t}  \leq 1-\lim\limits_{t \to \infty} \mathrm{erf}\left(\frac{k}{\sqrt{8 \lambda t}}\right).
\end{align}
\noindent Since Alice fingerprints the flows in the second phase, $t=T_2$. Recall that the probability that Alice buffers more than $\epsilon \sqrt{{ \lambda T_1 }/{m}}$ packets tends to one, as $T \to \infty$. Therefore, we let $k=\epsilon \sqrt{{ \lambda T_1 }/{m}}$. By~\eqref{eq:125}, the probability that Alice runs out of packets for flow $f_i^{(A)}$ satisfies:
\begin{align}
\label{eq:111}\lim\limits_{T \to \infty}  P_{\mathrm{f}_1}\leq 1-\lim\limits_{T \to \infty} \mathrm{erf}\left(\frac{\epsilon}{2}\sqrt{\frac{T_1}{2 m T_2}}\right)= 1- \mathrm{erf}\left({\epsilon}\sqrt{\frac{\alpha}{8 }}\right). 
\end{align}
\noindent where the equality holds since $T_1/T_2= m \alpha$ following from~\eqref{eq:lph10} and~\eqref{eq:lph20}. Note that~\eqref{eq:111} is independent of $i$ (index of the flow), and holds for all flows $f_i^{(A)}$, $1\leq i\leq m$. Let 
\begin{align}
\label{eq:alpha}\alpha&=(8/\epsilon^2)( \mathrm{erf}^{-1}(1-{\zeta}))^2.
\end{align}
By~\eqref{eq:alpha},~\eqref{eq:111} yields
\begin{align}
\label{eq:114}  P_{\mathrm{f}_1}\leq \zeta \text{ as } T \to \infty.
\end{align}
\noindent Consequently, by~\eqref{eq:uni},~\eqref{eq:112},~\eqref{eq:114}, $\mathbb{P}_f\leq \zeta$ for all $\zeta>0$, when $T \to \infty$ and thus Alice and Bob's fingerprinting is reliable. 

({\em Number of flows}) By~\eqref{eq:cap} and~\eqref{eq:lph20}, we require
\begin{align}
\label{eq:29}\frac{\log m}{T_2}  = \frac{(1+m \alpha)\log m}{T} < C.
\end{align}
\noindent as $T_2 \to \infty$ ($T \to \infty$). In Appendix~\ref{ap.0} we show that we can achieve~\eqref{eq:29} as long as 
\begin{align}
\label{eq:11} m= \frac{1}{2} \min\left\{ {\alpha}^{-1}\left({\frac{TC}{W(TC)}-1}\right),\frac{TC}{W(TC)}\right\},
\end{align}
\noindent where $W(\cdot)$ is the Lambert-W function. Since for $T>e$, $W(T)\leq \ln(T)$, Alice and Bob can invisibly and reliably pair the end points of every flow, and thus break the anonymity of a network (Setting 1 shown in Fig.~\ref{fig:SysModa}) with $m=\mathcal{O}(T/\log T)$ flows.  

\end{proof}
 
\section{Scenario~2: Alice fingerprint a subset of the flows, Setting 1} \label{scen2}

In Scenario~1, Willie is certain that if $H_1$ is true, i.e., Alice fingerprints, then all flows are slowed down in the first phase. In Scenario~2, we add uncertainty to Willie's knowledge under $H_1$: Alice fingerprints a subset $\mathcal{F}_f$ of the flows, and $\mathcal{F}_f$ is unknown to Willie. Therefore, Willie has to investigate a large set of flows to detect if some are slowed down in the first phase as required for fingerprinting. We show that Willie's uncertainty allows Alice to fingerprint more flows without being visible.

Alice fingerprints a subset of the flows she observes ($\mathcal{F}_{f} \subset \mathcal{F}_{A}$). For each flow, she selects a unique fingerprint from her codebook and alters the timings of that flow according it. Similar to Scenario~1, Alice has $T$ units of time which she divides into two phases: a buffering phase of length $T_1$, which ensures Alice can successfully fingerprint, and a fingerprinting phase of length $T_2=T-T_1$. Bob, who has access to the fingerprint codebook and observes the set of fingerprinted flows ($\mathcal{F}_{B}=\mathcal{F}_{f}$), extracts the fingerprints from the flows. The fingerprint codebook is secret and Willie does not have access to it. The network is modeled by $M$ parallel $M/M/1$ queues with each queue shared by a flow from Alice to Bob (main flow) as well as other interfering flows independent of the main flow (Setting~1 shown in Fig.~\ref{fig:SysModa}). We calculate the number of flows ($m$) that Alice can fingerprint using this scheme, asymptotically as a function of $T$.

\begin{mthm} {2}\label{thm:pflows} 
Consider Setting~1 (see Fig.~\ref{fig:SysModa}). In a set $\mathcal{F}_{A}$ containing $M$ flows with equal rates ($\lambda$), if Bob observes only the fingerprinted flows ($\mathcal{F}_B=\mathcal{F}_f$), Alice and Bob can invisibly and reliably trace $m$ flows in a time interval of length $T$, where 
\begin{align}
\label{eq:bigM}
m= \begin{cases} 
M, & M=\mathcal{O}(1) \\
o(\min\{\sqrt{M},e^{TC_1}\}), & M=\omega(1) \text{ \& } M=\mathcal{O}(e^{2 TC}) \\
\Theta(e^{TC_2}), & M=\omega(e^{2 TC}) 
\end{cases}
\end{align}
$C$ is given in~\eqref{eq:c}, and $C_1,C_2\in(0,C)$ are arbitrary constants. 
\end{mthm}
A more accurate characterization of $m$ with respect to $M$ is presented in~\eqref{eq:1} in the proof below.


\begin{proof}
\textbf{Construction}: The construction is similar to that of Scenario~1 except that Alice fingerprints a subset of the flows that she observes. Recall that all of flows observed by Bob are also observed by Alice ($\mathcal{F}_B \subset \mathcal{F}_A$). Alice knows which set of her flows will be observed by Bob, and chooses them for fingerprinting ($\mathcal{F}_f= \mathcal{F}_B$). Note that Willie does not know which subset of $\mathcal{F}_A$ is $\mathcal{F}_B$. Alice generates a codebook of $m$ fingerprints (similar to Scenario~1) and shares it with Bob prior to fingerprinting, where $m$ is given in~\eqref{eq:bigM}. Recall that we calculate the maximum number of flows that Alice and Bob can trace; therefore, we only consider the case $|\mathcal{F}_B|=m$ which can be extended to $|\mathcal{F}_B|\leq m$ trivially.

Alice's scheme consists of two phases, a buffering phase of length $T_1$, and a fingerprinting phase of length $T_2=T-T_1$, where
\begin{align}
\label{eq:lph1} T_1&=\frac{T \alpha'}{\ln(1+{\frac{\epsilon^2 M }{2 m^2}})+\alpha'},\\
\label{eq:lph2} T_2&=T-T_1=\frac{T }{1+\alpha'/\ln(1+{\frac{\epsilon^2 M }{2 m^2}})},\\
\label{eq:alphap} \alpha'&=\alpha \epsilon^2.
\end{align}
Recall that $\alpha$ and $C$ are given in~\eqref{eq:alpha} and~\eqref{eq:c}, respectively, and $\epsilon$ is the invisibility parameter. Alice generates fingerprints for her codebook analogous to Scenario~1. The number of fingerprints in her codebook is $m$.

\textbf{Analysis}: ({\em Invisibility}) For each phase, we show that all operations Alice performs on the flows are invisible. Consider the first phase $[0,T_1]$ where Alice slows down each flow from rate $\lambda$ to rate $\lambda-\Delta$ with
\begin{align}
\label{eq:del2}\Delta = \sqrt{\frac{\lambda}{T_1} \ln\left(1+{\frac{\epsilon^2 M }{2 m^2}}\right)}.
\end{align}
\noindent From Willie's perspective, the number packets in time $[0,T_1]$ is a sufficient statistic to detect Alice~\cite{soltani2015covert}. If Alice does not fingerprint ($H_0$), then the joint probability density function (pdf) of Willie's observations is $\mathbb{P}_0= \prod_{i=1}^{M} \mathbb{P}_{\lambda}(n_i)$ where $\mathbb{P}_{\lambda}(n)$ is the pdf of a Poisson random variable with mean $\lambda$. Note that Willie knows that $m$ out of $M$ flows observed by Alice is selected to be fingerprinted, but he does not know which set is selected. Therefore, from Willie's point of view, if Alice chooses to fingerprint flows ($H_1$), then each flow will contain a fingerprint with probability  
\begin{align}
\label{eq:p1} p=\frac{m}{M}.
\end{align}
Thus, the joint pdfs of Willie's observations when Alice fingerprints ($H_1$) is
\begin{align}
\nonumber \mathbb{P}_1&=\prod_{i=1}^{M} \left(p\mathbb{P}_{\lambda-\Delta}(n_i)+(1-p)\mathbb{P}_{\lambda}(n_i)\right),
\end{align}
\noindent where $\Delta$ is the change in flow rate. Note that the change of rate differs from the one in Scenario~1. Suppose that Willie applies an optimal hypothesis test to minimize his probability of error $\mathbb{P}_{\mathrm e}^{(\mathrm w)}$. Then, we can obtain a lower bound on his probability of error\cite[Eq.1]{soltani2018covert}:
\begin{align} 
\label{eq:0} \mathbb{P}_{\mathrm e}^{(\mathrm w)} \geq \frac{1}{2}- \sqrt{\frac{1}{8} \mathcal{D}(\mathbb{P}_1 || \mathbb{P}_0)},
\end{align}
\noindent where $\mathcal{D}(\mathbb{P}_1 || \mathbb{P}_0)$ is the Kullback–Leibler divergence (relative entropy) between $\mathbb{P}_1$ and  $\mathbb{P}_0$.

Alice's scheme is invisible as long as she can make Willie's detector operate as close as desired to the detector that disregards Willie's observations and results in $\mathbb{P}_{\mathrm e}^{(\mathrm w)}=1/2$ (see Definition~\ref{def:inv}). In Appendix~\ref{ap.4}, we show that for $\epsilon>0$, 
\begin{align}
\label{eq:1234}\sqrt{\frac{1}{8} \mathcal{D}(\mathbb{P}_1 || \mathbb{P}_0)}\leq \epsilon.
\end{align}
\noindent Thus,~\eqref{eq:0} yields $\mathbb{P}_{\mathrm e}^{(\mathrm w)} \geq \frac{1}{2}-\epsilon$ as $T \to \infty$, and thus Alice's buffering is invisible.

The second phase is invisible because the fingerprints are samples of Poisson processes with rate $\lambda$. Combined with the invisibility of the first phase, Alice and Bob's scheme is invisible. 

({\em Reliability}) The analysis is similar to that of Scenario~1. Since all flows observed by Bob are fingerprinted ($\mathbb{P}_{\mathrm{f}_3}=0$), to show Alice and Bob's scheme is reliable, it suffices to show that for each flow $\mathbb{P}_{\mathrm{f}_1}+\mathbb{P}_{\mathrm{f}_2}\leq \zeta$ for all $\zeta>0$.

Similar to Scenario~1, in Appendix~\ref{ap.5} we show that   
\begin{align}
\label{eq:4} P_{\mathrm{f}_1}\leq \zeta \text{ as } T \to \infty.
\end{align}
 
Now, consider Bob's decoding error for each flow, $\mathbb{P}_{\mathrm{f}_2}$. By~\eqref{eq:lph1} and~\eqref{eq:lph2}, $T_1,T_2\to \infty$ as $T \to \infty$. In order for Bob to be able to successfully extract the fingerprint from each flow, we require
\begin{align}
\label{eq:cap2} \frac{\log{m} }{T_2}&< C. 
\end{align}
\noindent as $T_2 \to \infty$ ($T \to \infty$). Substituting $T_2$ from~\eqref{eq:lph2} and re-arranging yields:
\begin{align}
\label{eq:1} m \leq \exp\left( \frac{TC }{1+\alpha'/\ln(1+{\frac{\epsilon^2 M }{2 m^2}})}\right)
\end{align}
We show in Appendix~\ref{ap.3} that~\eqref{eq:1} holds asymptotically as $T \to \infty$, given the value of $m$ provided in~\eqref{eq:bigM}.

Consequently,
\begin{align}
\label{eq:5}\mathbb{P}_{\mathrm{f}_2} \to 0 \text{ as } T \to \infty.
\end{align}
\noindent By~\eqref{eq:uni},~\eqref{eq:4}, and~\eqref{eq:5}, $\mathbb{P}_f \to 0$ as $T \to \infty$. Thus, if $M=\omega(1)$, Alice can invisibly and reliably fingerprint $o\left(\min\{\sqrt{M},e^{T C}\}\right)$ flows in a time interval of length $T$, and Bob can successfully extract the fingerprints,  
where $C$ is given in~\eqref{eq:c}, and if $M=\mathcal{O}(1)$, Alice can invisibly and reliably fingerprint all $M$ flows in a time interval of length $T$, and Bob can successfully extract the fingerprints.
\end{proof}

In Scenario~2, we assumed that all flows observed by Bob are also observed by Alice and chosen for fingerprinting ($\mathcal{F}_f=\mathcal{F}_B\subset \mathcal{F}_A$). Although this is applicable in many schemes, we present results for the case where this assumption is relaxed in Section~\ref{sec4}, i.e., Bob observes flows with and without fingerprints.

\section{Extension to arbitrary rates} \label{sec3}
In this section, we extend Theorems~\ref{thm:allflows} and~\ref{thm:pflows} to the case that the flow rates are arbitrary. 

\subsection{Scenario 3: All flows are fingerprinted and flow rates are arbitrary, Setting 1} \label{scen3}

Consider Scenario~3, which is the extension of Scenario~1 to arbitrary rates: Alice fingerprints all of the flows she observes ($\mathcal{F}_{f}=\mathcal{F}_{A}$), and Bob observes only the fingerprinted flows ($\mathcal{F}_{B}=\mathcal{F}_f$). We consider Setting 1 (see Fig.~\ref{fig:SysModa}), i.e., $M$ parallel $M/M/1$ queues with multiple inputs and outputs, where each queue is shared between a flow from Alice to Bob (main flow) as well as other interfering flows independent of the main flow. Here the flows rates $\lambda_1,\ldots,\lambda_M$ can be arbitrary, and the main flow passing through the $i^{\mathrm{th}}$ queue ($q_i$) has the rate of $\lambda_i$. Alice fingerprints the input flows of the network in the time interval $[0,T]$, and Bob extracts the fingerprints from the flows on the output links of the network to infer the connections between input and output flows. 

Similar to Scenario~1, for each flow Alice selects a codeword (fingerprint) from her codebook and embeds it in the flow by changing the packet timings of the flow. She builds her codebook based on the minimum rate of the flows $\lambda_{\mathrm{min}}=\min(\lambda_1,\ldots,\lambda_M)$, and to embed a fingerprint (of rate $\lambda_{\mathrm{min}}$) in a flow of rate $\lambda_i$, she scales the fingerprint by a factor of $\lambda_{\mathrm{min}}/\lambda_i$ to obtain a modified fingerprint of rate $\lambda_i$, and then embeds it in the flow. In addition, she uses a two-phase (buffering-fingerprinting) scheme similar to those of Scenarios~1 and~2. 

We calculate the number of flows ($m=M$) that Alice and Bob can trace by fingerprinting using this scheme, asymptotically as a function of $T$.
\begin{mthm}{3.1} \label{thm:allflowsd} 
	Consider Setting~1 (see Fig.~\ref{fig:SysModa}). If Alice fingerprints all $M$ input flows ($\mathcal{F}_{f}=\mathcal{F}_{A}$) whose rates $\lambda_1,\ldots,\lambda_m$ are arbitrary and Bob observes only the set of fingerprinted flows ($\mathcal{F}_{B}=\mathcal{F}_{f}$), then Alice and Bob can invisibly and reliably trace $m=M=\mathcal{O}(T/\log T)$ flows in a time interval of length $T$.
\end{mthm}

\begin{proof}
	\textbf{Construction}: 
	Per above, Alice employs a two-phase scheme: a buffering phase of length $T_1$ and a fingerprinting phase of length $T_2=T-T_1$ (see Fig.~\ref{fig:Twophased}), where $T_1$ and $T_2$ are given in ~\eqref{eq:lph10} and~\eqref{eq:lph20}. The codebook construction is similar to Scenario~1, but the rate of the fingerprints (codewords) is $\lambda_{\mathrm{min}}=\min(\lambda_1,\ldots,\lambda_M)$. To embed a fingerprint in a flow of rate $\lambda_i$, Alice selects a fingerprint $(\tau_1,\ldots,\tau_N)$ and scales by a factor ${\lambda_{\mathrm{min}}}/\lambda_i$ to generate a modified fingerprint of rate $\lambda_i$, $(\frac{\lambda_{\mathrm{min}} \tau_1}{\lambda_i} ,\ldots,\frac{\lambda_{\mathrm{min}} \tau_N}{\lambda_i})$. Since fingerprints are instantiations of a Poisson process of parameter $\lambda_{\mathrm{min}}$ (i.e., its inter-arrival times are instantiations of an exponential random variable of mean $1/\lambda_{\mathrm{min}}$), the modified fingerprint is an instantiation of a Poisson process of parameters $\lambda_i$.
	Next, Alice applies the inter-packet delays given by the modified fingerprint to each flow. 
	
	
	Recall that Bob knows the rate of each flow. Upon observing $f_{i}^{(B)}$, the flow with packet timings $\bar{t_i}=(t_i^{(1)},t_i^{(2)},\ldots,t_i^{(N)})$ and rate $\lambda_i$, Bob seeks to answer the following question:
	
	\indent \textbf{Question~1:}\textit{ Given that Alice used the codebook $\{C(W_l)\}_{l=1}^{l=m}$ whose fingerprints are of rate $\lambda_{\min}$, what is the index of the fingerprint that was selected by Alice, scaled to rate $\lambda_i$, and transmitted through $q_i$ to produce the output packet timings $\bar{t_i}$?}

\textbf{Analysis}: ({\em Invisibility}) Similar to Scenario~1, we analyze the invisibility of the first and second phases separately. In the first phase, Alice slows down each flow of rate $\lambda_i$ to rate $\lambda_i-\epsilon \sqrt{{2 \lambda_i}/{m T_1}}$. Using arguments similar to that of Theorem~\ref{thm:allflows}, we can show that~\cite[Theorem 2]{soltani2015covert}:
\begin{align}
\nonumber \mathbb{P}_{\mathrm e}^{(\mathrm w)}>\frac{1}{2}-{\epsilon},
\end{align}
\noindent where $\mathbb{P}_{\mathrm e}^{(w)}$ is Willie's error probability. Thus, this phase is invisible to Willie. In the second phase, since Alice embeds a modified fingerprint of rate $\lambda_i$ in a flow of rate $\lambda_i$, the traffic pattern remains Poisson with rate $\lambda_i$ indistinguishable from the pattern that Willie expects to observe. Hence, the scheme is invisible.

({\em Reliability}) Similar to the reliability analysis in Scenario~1, we upper bound $\mathbb{P}_{\mathrm{f}_1}+\mathbb{P}_{\mathrm{f}_2}$ by $\zeta$, for all $\zeta>0$.

Recall that upon observing $f_{i}^{(B)}$, Bob seeks the answer to Question~1. Note that the answer to this question is the same as the answer to the following question:

\indent \textbf{Question~2: }\textit{Given that Alice used the codebook $\{C'(W_l)\}_{l=1}^{l=m}=\frac{\lambda_{\min}}{\lambda_{i}}\{C(W_l)\}_{l=1}^{l=m}$ what is the index of the fingerprint that was selected by Alice and transmitted through $q_i$ to produce the output packet timings $\bar{t_i}$?}

In other words, although Alice generates a codebook whose fingerprints are of rate $\lambda_{\min}$ and then scales each fingerprint to adjust to rate $\lambda_i$ of the flow, Bob's decoding of each flow is equivalent to the case where Alice uses a codebook whose fingerprints are of rate $\lambda_i$ and she does not scale the fingerprints; the only differences are in the number of fingerprints (codewords) and the time to transmit the fingerprint, as we will explain later. Therefore, from~\eqref{eq:22}, Bob can successfully extract the fingerprint from the flow of rate $\lambda_i$ as long as $T_2$ is large and
\begin{align}
\label{eq:221} \frac{\log m}{T_2^{(i)}}< \lambda_i \log{\left({(\mu_i-\lambda'_i)}/{\lambda_i}\right)},
\end{align}
\noindent where $T_2^{(i)}=T_2 \lambda_{\min}/\lambda_i$ is the time of the transmission of the fingerprint embedded in the flow of rate $\lambda_i$. Therefore, 
\begin{align}
\label{eq:222}  \frac{\log m}{T_2 }<\lambda_{\min}  \log{\left({(\mu_i-\lambda'_i)}/{\lambda_i}\right)}.
\end{align}
\noindent Since the size of the codebook is $m$, fingerprinting the flow $f_i$ corresponds to transmission of $\log{m}$ nats of information through the inter-packet delays of the flow $f_i$. Note that scaling a fingerprint of rate $\lambda_{\min}$ to rate $\lambda_i$ results in transmission of $\log{m}$ nats of information at a higher rate but a shorter time.

Since~\eqref{eq:222} holds for all $1\leq i\leq m$, we require
\begin{align}
\label{eq:cap3} \frac{\log m}{T_2}< C',
\end{align}
\noindent where 
\begin{align}
\label{eq:c2} C'=\lambda_{\min} \underset{i}{\mathrm{min}}\{\log{\left({(\mu_i-\lambda'_i)}/{\lambda_i}\right)}\},
\end{align}
\noindent to achieve $\mathbb{P}_{\mathrm{f}_2} \to 0 \text{ as } T_2 \to \infty$ for each flow. Note that~\eqref{eq:lph20} implies that $T_1,T_2 \to \infty$ as $T \to \infty$. Therefore,
\begin{align}
\label{eq:1121}\mathbb{P}_{\mathrm{f}_2} \to 0 \text{ as } T \to \infty.
\end{align}

Now, consider $\mathbb{P}_{\mathrm{f}_1}$. In the second phase, on each link Alice receives and transmits the packets according to two independent Poisson processes of equal rate. Thus, we employ a random walk analysis similar to that of Scenario~1 to show that 
\begin{align}
\label{eq:1141} P_{\mathrm{f}_1}\leq 1-\mathrm{erf}\left(\frac{\epsilon}{2}\sqrt{\frac{T_1}{2 m T_2}}\right) \leq \zeta \text{ as } T \to \infty.
\end{align}
\noindent Consequently, by~\eqref{eq:uni},~\eqref{eq:1121} and~\eqref{eq:1141}, $\mathbb{P}_f\leq \zeta$ for all $\zeta>0$, and thus Alice and Bob's fingerprinting is reliable. 

({\em Number of flows}) The analysis is similar to that of Scenario~1. As $T \to \infty$, we require
\begin{align}
\label{eq:291}\frac{\log m}{T_2}  = \frac{(1+m \alpha)\log m}{T} < C'.
\end{align}
\noindent which we can achieve as long as 
\begin{align}
\label{eq:1111} m= \frac{1}{2} \min\left\{ {\alpha}^{-1}\left({\frac{TC'}{W(TC')}-1}\right),\frac{TC'}{W(TC')}\right\},
\end{align}
Since for $T>e$, $W(T)\leq \ln(T)$, Alice and Bob can invisibly and reliably break the anonymity of a network (Setting 1 shown in Fig.~\ref{fig:SysModa}) with  $m=\mathcal{O}(T/\log T)$ flows. 
\end{proof}

\subsection{Scenario 4: Alice fingerprints a subset of the flows, Setting 1} \label{scen4}
Consider Scenario~4, which is the extension of Scenario~2 to arbitrary rates: Alice fingerprints a subset $\mathcal{F}_f$ of the flows, and $\mathcal{F}_f$ is unknown to Willie. Similar to Scenario~2, since Willie has to investigate a large set of flows to detect if some are slowed down in the first phase as required for fingerprinting, Alice can make more fingerprinted flows invisible.

For each flow in $\mathcal{F}_f$, she selects a unique fingerprint from her codebook and alters the timings of that flow according to the fingerprint. We consider Setting 1 (see Fig.~\ref{fig:SysModa}), i.e., $M$ parallel $M/M/1$ queues with multiple inputs and outputs, where each queue is shared between a flow from Alice to Bob (main flow) as well as other interfering flows independent of the main flow. Flows rates are $\lambda_1,\ldots,\lambda_M$, which can be arbitrary, and the main flow passing through the $i^{\mathrm{th}}$ queue ($q_i$) has the rate of $\lambda_i$. Alice fingerprints the input flows of the network in the time interval $[0,T]$, and Bob extracts the fingerprints from the flows on the output links of the network to infer the connections between input and output flows. 

For each selected flow Alice selects a codeword from her codebook and embeds it in the flow by changing its packet timings according to the selected fingerprint. Since flow rates are arbitrary, similar to Scenario~3, she builds her codebook based on the minimum rate of the flows to be fingerprinted and scales each fingerprint based on the rate of the flow to be fingerprinted. Also, she uses a two-phase (buffering-fingerprinting) scheme.

We calculate the number of flows ($m$) in which Alice fingerprints using this scheme, asymptotically as a function of $T$.
\begin{mthm}{3.2} \label{thm:pflowsd} Consider Setting~1 (see Fig.~\ref{fig:SysModa}). In a set $\mathcal{F}_{A}$ containing $M$ flows with rates $\lambda_1,\ldots,\lambda_M$, if Bob observes only the fingerprinted flows ($\mathcal{F}_B=\mathcal{F}_f$), Alice and Bob can invisibly and reliably trace $m$ flows in a time interval of length $T$, where $m$ is given in~\eqref{eq:bigM}, where $C$ is replaced with $C'$ which is given in~\eqref{eq:c2}.
\end{mthm}

\begin{proof}
		The construction and analysis follow from those of Scenarios~2 with modifications due to arbitrary rates. The extension to arbitrary rates follows from that of Scenario~3. 
\end{proof}

\section{Mixing flows with and without fingerprints} \label{sec4}
We have previously assumed that Bob only observes the set of fingerprinted flows, i.e., $\mathcal{F}_{B}=\mathcal{F}_f$. But, in practice Bob might observe a set of flows in which some of the flows are not fingerprinted, and therefore, he must be able to detect if a flow contains a fingerprint. In this Section, we consider Setting 2 (see Fig.~\ref{fig:SysModb}) and we present Scenarios~5 and~6 which are extensions of Scenarios~1 and~2, respectively, to the case where Bob observes a set of flows in which some of them are not fingerprinted. We present a detector for Bob that is able to detect if a flow is fingerprinted. 
\subsection{Scenario 5: All flows are fingerprinted and Bob observes flows with and without fingerprints, Setting 2} \label{scen5}
Consider Scenario~5, which is the extension of Scenario~1 to the case where Bob observes flows with and without fingerprints ($\mathcal{F}_f \subset \mathcal{F}_{B}$): Alice fingerprints all of the flows she observes ($\mathcal{F}_{f}=\mathcal{F}_{A}$), flow rates are equal ($\lambda$), and Bob observes flows with and without fingerprints. We consider Setting 2 (see Fig.~\ref{fig:SysModb}), i.e., $M$ parallel $M/M/1$ queues with single input and output. Alice fingerprints the input flows of the network in the time interval $[0,T]$, and Bob extracts the fingerprints from the flows on the output links of the network to infer the connections between input and output flows. 

In contrast to Scenarios~1-4, Bob uses a detector to determine if a flow is fingerprinted. We calculate the number of flows ($m$) that Alice and Bob can trace by fingerprinting using this scheme, asymptotically as a function of $T$.

\begin{mthm}{4.1} \label{thm:allflowsw} 
	Consider Setting~2 (see Fig.~\ref{fig:SysModb}). If Alice fingerprints all $M$ input flows ($\mathcal{F}_{f}=\mathcal{F}_{A}$) whose rates are equal ($\lambda$) and Bob observes a set of flows with and without fingerprints ($\mathcal{F}_{f}\subset\mathcal{F}_{B}$), then Alice and Bob can invisibly and reliably trace $m=M=\mathcal{O}(T/\log T)$ flows in a time interval of length $T$.
\end{mthm}

\begin{proof}
\textbf{Construction}: The only difference between the construction of Scenarios~1 and~5 is that, for Scenario~5, Bob must use a detector which detects if a flow contains a fingerprint. 
	
Here, Bob's decoder is different from the maximum likelihood decoder proposed in~\cite[p. 9]{verdubitsq}, which for each codeword calculates the service times that yield $\bar{D_i}$, removes the codewords that result in negative values of service times, and finally finds a unique codeword that corresponds to the minimum sum of service times. Instead, Bob's decoder selects a threshold $\beta=\log{(\mu_i/\lambda)}$, applies a function on each codeword, and finds a unique codeword that generates an output for the function that is larger than $\beta$. 
	
Next, we describe Bob's decoder in detail~\cite[p. 12]{sundaresan2000robust}. For $\bar{x}=(x_1,\ldots,x_n)\in \mathcal{R}_{+}^{n}$ and $\bar{y}=(y_0,y_1,\ldots,y_n)\in \mathcal{R}_{+}^{n+1}$, if $\bar{x}$ is the sequence of packet timings before the flow passes through $q_i$ (inter-arrival times), then the pdf of the observed packet timings $\bar{y}$ (inter-departure times) is:
\begin{align}
\nonumber \mathbb{P}(\bar{y}|\bar{x}) = e_{\mu-\lambda} (y_0)\prod_{k=1}^{n}(y_k-w_k),
\end{align}
\noindent where $e_{u}(x)=u e^{-u x}$ is the exponential pdf with mean $1/u$, and $w_k = \max\{0,\sum_{i=1}^{k}x_i-\sum_{i=0}^{k-1}y_i\}$ is the $k^{\mathrm{th}}$ waiting time, the amount of time that the queue waits until it receives the $k^{th}$ packet. Since the packet timings of the fingerprinted flow is an instantiation of a Poisson process of rate $\lambda$, the joint pdf of the inter-arrival times is $\prod_{k=1}^{n} e_{\lambda}(x_k)$. Consequently, the pdf of $\bar{y}$ is:
\begin{align}
	\mathbb{P}(\bar{y}) &= \int_{\mathbb{R}_{+}^{n}} \mathbb{P}(\bar{y}|\bar{x}) \prod_{k=1}^{n} e_{\lambda}(x_k) d\bar{x}.
\end{align}
\noindent Bob's decoder finds a unique fingerprint (codeword) $W_l$ from $\{W_l\}_{l=1}^{l=m}$ that satisfies $\frac{\mathbb{P}(\bar{y}|W_l)}{\mathbb{P}(\bar{y})} >\beta$; if such  a unique codeword does not exist, it outputs \textit{flow not fingerprinted}. 

\textbf{Analysis}: The analysis follows from that of Scenario~1. The only differences appear in the analysis of Bob's decoding error probability. The auxiliary threshold decoder used in the analysis of the mismatched decoder in~\cite[p. 413-417]{sundaresan2000robust} provides what we need for our application. If Bob uses this detector, the decoding error probability of a fingerprinted flow will be:
\begin{align}
\label{eq:121}\mathbb{P}_{\mathrm{f}_2} \to 0 \text{ as }  T \to \infty,
\end{align}
\noindent which implies that if we generate $m$ independent instantiations of a Poisson process of rate $\lambda$ on a time interval of length $T_2$, $W_1,\ldots,W_m$, we select one of them $W_l$ and send a packet stream whose packet timings follow $W_l$ over the network, then the probability that at least one $W_k\neq W_l$ satisfies $\frac{\mathbb{P}(\bar{y}|W_l)}{\mathbb{P}(\bar{y})} >\beta$ tends to zero, i.e., 
\begin{align}
\label{eq:122} \mathbb{P}(\exists W_{k \neq l}:{\mathbb{P}(\bar{y}|W_k)}/{\mathbb{P}(\bar{y})} >\beta |W_l \text{ sent }) \to 0 \text{ as } T\to\infty. 
\end{align}
\noindent Consider the case where Bob observes a flow that is not fingerprinted. Recall that the packet timings of all the flows follow a Poisson process of rate $\lambda$. Denote by $Z^{\star}$ an instantiation of a Poisson process that corresponds to the packet timings of the this flow before it passes through the network. If Bob detects a fingerprint, it must be that one of the fingerprints $W_{l}$ in the codebook resulted in $\frac{\mathbb{P}(\bar{y}|W_l)}{\mathbb{P}(\bar{y})} >\beta$. Hence, 
\begin{align}
\mathbb{P}_{\mathrm{f}_3}  = \mathbb{P}\left(\exists W_{k}:\frac{\mathbb{P}(\bar{y}|W_k)}{\mathbb{P}(\bar{y})} >\beta \bigg|Z^{\star} \text{ sent }\right) 
\end{align}	
\noindent Recalling that $W_1,\ldots,W_m$ and $Z^{\star}$ are independent instantiations of a Poisson process of rate $\lambda$,~\eqref{eq:121} and~\eqref{eq:122} yield $\mathbb{P}_{\mathrm{f}_3} \to 0$ as $T \to \infty$. Thus, Alice and Bob's fingerprinting is reliable. 
\end{proof}
\subsection{Scenario 6: Alice fingerprints a subset of the flows and Bob observes flows with and without fingerprints, Setting 2} \label{scen6}

Consider Scenario~6, which is the extension of Scenario~2 to the case where Bob observes flows with and without fingerprints ($\mathcal{F}_f \subset \mathcal{F}_{B}$): Alice fingerprints a subset of the flows she observes ($\mathcal{F}_{f}\subset \mathcal{F}_{A}$), flow rates are equal ($\lambda$), and Bob observes flows with and without fingerprints. We consider Setting 2 (see Fig.~\ref{fig:SysModb}), i.e., $M$ parallel $M/M/1$ queues with single input and output. Alice fingerprints the input flows of the network in the time interval $[0,T]$, and Bob extracts the fingerprints from the flows on the output links of the network to infer the connections between input and output flows. 

Similar to Scenario~5, Bob's detector is able to distinguish whether a flow is fingerprinted. We calculate the number of flows ($m$) that Alice and Bob can trace by fingerprinting, asymptotically as a function of $T$.
\begin{mthm} {4.2}\label{thm:pflowsw} 
Consider Setting~2 (see Fig.~\ref{fig:SysModb}). In a set $\mathcal{F}_{A}$ containing $M$ flows with equal rates ($\lambda$), if Bob observes flows with and without fingerprints ($\mathcal{F}_f\subset\mathcal{F}_B$), Alice and Bob can invisibly and reliably trace $m$ flows in a time interval of length $T$, where $m$ is given in~\eqref{eq:bigM}, where $C$ is replaced with
\begin{align}
\label{eq:c3} C''=\lambda \log{\left({\underset{i}{\mathrm{min}} \{\mu_i\}}/{\lambda}\right)},
\end{align}
\end{mthm}
Note that the replacement of $C$ with $C''$ is necessary since here we consider Setting~1 which implies that the rates of interfering flows $\lambda_i'$ are zero.
\begin{proof}
	The construction and analysis follow from those of Scenarios~2 with modifications due to the change of Bob's detector to detect whether a flow is fingerprinted or not. In addition, Alice does not need to know which subset of the flows she observes are observed by Bob to fingerprint the. But, she chooses an arbitrary subset of flows and fingerprints them. In general, each fingerprinted flow will not be observed by Bob. However, since we determine the maximum number of flows that can be traced, we assume that each fingerprinted flow will be observed by Alice. The analysis for Bob's detector follows from that of Scenario~5. 
\end{proof}

In Theorem~\ref{thm:pflowsw}, Alice's selection of subset might be due to the preference of Alice and Bob. But, if there is no such preference, Alice can choose the flows randomly and independently to fingerprint them. Next, we present Theorem~\ref{thm:pflowsw2} to address this case.

\begin{mthm} {4.3}\label{thm:pflowsw2} 
	Consider Setting~2 (see Fig.~\ref{fig:SysModb}). In a set $\mathcal{F}_A$ containing $M$ flows, if Alice fingerprints each flow independently with probability $q$, each flow has rate $\lambda$, and Bob observes a set of flows that contains flows with and without fingerprints ($\mathcal{F}_{f}\subset \mathcal{F}_{B}$), then Alice and Bob can invisibly and reliably trace 
	\begin{align} \label{eq:120}
m=O\left(\min\left\{M q, \exp\left( \frac{TC }{1+\alpha'/\ln(1+{\frac{\epsilon^2  }{2 M q^2}})}\right) \right\}\right)	
	\end{align}
	flows in a time interval of length $T$, where $\epsilon$ is the invisibility parameter, and $C''$ and $\alpha'$ are given in~\eqref{eq:alphap} and~\eqref{eq:c3}, respectively.
\end{mthm}

\begin{proof}
	The construction and analysis follows those of Theorem~\ref{thm:pflowsw} with modifications due to the random selection of the flows. Alice builds a fingerprint codebook of size $m$, where
	\begin{align}
	\label{eq:14} m = \exp\left( \frac{TC }{1+\alpha'/\ln(1+{\frac{\epsilon^2 }{2 M q^2}})}\right).
	\end{align}
	She selects the flow $f_i^{(A)}\in \mathcal{F}_{A}$ to be fingerprinted with probability $q$, independent of other flows. For each flow $f_i^{(A)}$ she generates an independent Bernoulli random variable $X_i$ with $P(X_i=1)=q$; she selects a unique (unused) fingerprint from her codebook and embeds it in flow $f_i^{(A)}$ if and only if $X_i=1$. 
	
	Similar to the analysis of Scenario~2, we can show that for reliable fingerprinting we require 
	\begin{align}
	\label{eq:12} m \leq \exp\left( \frac{TC }{1+\alpha'/\ln(1+{\frac{\epsilon^2  }{2 M q^2}})}\right),
	\end{align}
	which is satisfied by~\eqref{eq:14}. 
	Next, we show that $N_s=\sum_{k=1}^{M} X_i = \mathcal{O}(Mq)$. Consider random variables $Y_i=X_i/q$, $i=1,\ldots, M$. Since $\mathbb{E}[Y_i]=\mathbb{E}[X_i]/q=1$, the weak law of large numbers (WLLN) yields $\lim\limits_{T \to \infty} \mathbb{P}\left(\frac{1}{M}\sum_{i=1}^{M}Y_i>1/2\right) = 1$. Let $\gamma=1/2$ and $X_i=q Y_i $. Thus, $\lim\limits_{T \to \infty} \mathbb{P}\left(\sum_{i=1}^{M}X_i>Mq/2\right) = 1$. Since Alice fingerprints $\min\{m,N_s\}$ flows, the number of flows that Alice and Bob can invisible and reliably trace is~\eqref{eq:120}.
	
In~\cite[Theorem 2]{soltani2017towards}, we presented values of $q$ and $M$ that yield a close to a maximal number of flows that can be traced.
\end{proof}

\section{Simulation Results}
\subsection{Willie's error probability}
First, we consider Scenario 1 and present the results of the simulation for Willie's detection. Then, we discuss how similar results apply to all of the scenarios with slight modifications. 

Consider Scenario 1. Recall that when $H_0$ is true (Alice is not fingerprinting), Willie observes flows where each flow's packet timing is governed by a Poisson process of rate $\lambda$. When $H_1$ is true (Alice is fingerprinting), the packet timing of each flow observed by Willie is governed by a Poisson process of rate $\lambda-\epsilon\sqrt{2\lambda/mT_1}$ in the first phase and a Poisson process of rate $\lambda$ in the second phase. Since the statistical properties of the flows are the same for $H_0$ and $H_1$ in the second phase, he uses the information obtained from his observations in the first phase to test whether Alice is fingerprinting. Note that when $H_1$ is true, Willie observes $m$ flows each of whose packet rates is $\lambda-\epsilon\sqrt{2\lambda/mT_1}$ in the first phase. Similar to~\cite{soltani2015covert}, we can show that a packet counter is an optimal detector for Willie. He counts the total number of packets $S$ in the first phase for all $m$ flows, and sets a threshold $U$. If $S<\lambda T m - U$, he selects $H_1$; otherwise, he selects $H_0$. 

\begin{figure}
	\begin{center}
		\includegraphics[width=\textwidth,height=\textheight,keepaspectratio]{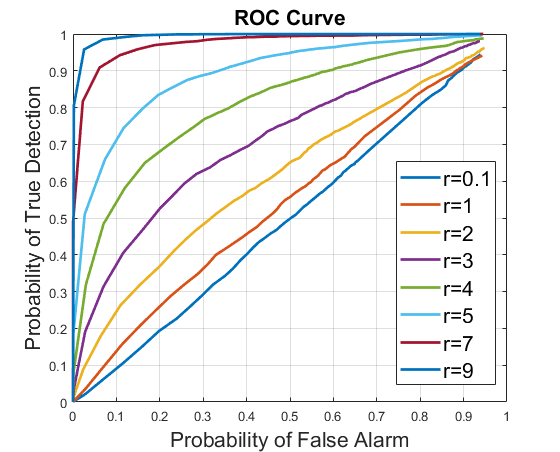}
	\end{center}
	\caption{The receiver operating characteristic (ROC) curve for Willie's detection. Alice reduces the rate of each packet stream from $\lambda$ to $\lambda-r \epsilon\sqrt{2\lambda/mT_1}$, and we draw ROC curves for $r\in [0.1,9]$. }
	\label{fig:roc}
\end{figure}

We consider $\mathbb{P}(H_0)=\mathbb{P}(H_1)=0.5$. The simulation parameters are $\lambda=7.36$ packets/second, ${\underset{i}{\mathrm{min}} \{\mu_i-\lambda'_i\}}=20$ packets/second, $C=7.36$  nats/second (see~\eqref{eq:c}), $T=3600 \times 11$ seconds, $\zeta=0.01$, $\epsilon=0.1$, $m=10$ (see~\eqref{eq:11}), $U\in [0.07,100] \sqrt{\lambda m T} \approx [80, 120000]$. Alice reduces the rate of each packet stream from $\lambda$ to $\lambda-r \epsilon\sqrt{2\lambda/mT_1}$, and we plot receiver operating characteristic (ROC) curves for Willie for $r\in [0.1,9]$ (see Fig.~\ref{fig:roc}). Note that the x-axis and y-axis of this figure are Willie's probability of false-alarm ($\mathbb{P}_{\mathrm{FA}}$) and true-detection (1-$\mathbb{P}_{\mathrm{MD}}$), respectively. The number of trials is $8000$. According to Theorem~\ref{thm:allflows}, $r=1$ corresponds to the case which yields covertness, as verified by the ROC curve. Note that large values for $r$, which corresponds to more slow down of the packets by Alice in the first phase,  lead to detection by Willie with high probability. 

Next, we discuss why these results apply to other scenarios. Note that Willie's detection defers across scenarios since he observes a different number of flows. However, in all scenarios Willie's optimal detector is a packet counter. Since all of the links are governed by independent Poisson processes and the sum of independent Poisson random variables (with distinct parameters) is another Poisson random variable, Willie's detection problem differs only slightly.

\subsection{Probability that Alice runs out of packets} \label{sec:pa}
Recall that in all scenarios Alice slightly slows down the packet rate of each flow so as to buffer packets. She does this to ensure that in the second phase, she does not run out of packets with high probability. We denoted the probability that Alice runs out of packets by $\mathbb{P}_{\mathrm{f}_1}$, and recall we want to achieve $\mathbb{P}_{\mathrm{f}_1}<\zeta$. 

\begin{figure}
	\begin{center}
		\includegraphics[width=\textwidth,height=\textheight,keepaspectratio]{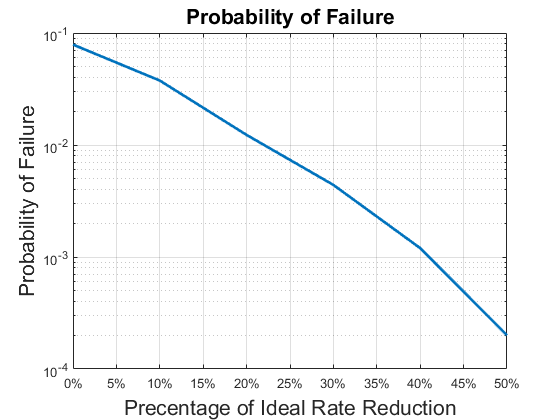}
	\end{center}
	\caption{The probability that Alice runs out of packets for a single link when, in the buffering phase, she reduces the packet rate from $\lambda$ to $\lambda -r' \epsilon \sqrt{2\lambda/(mT_1)} $, where $100 r'$ is the percentage of ideal rate reduction.}
	\label{fig:pf}
\end{figure}

We consider a single link and plot the curve for the probability that Alice runs out of packets when she reduces the rate from $\lambda$ to $\lambda -r' \epsilon \sqrt{2\lambda/(mT_1)} $ (see Fig.~\ref{fig:pf}), where $r'\in[0,0.5]$ is variable. We term $100 r'$ the percentage of ideal rate reduction. According to Theorem~\ref{thm:allflows}, $r'=1$ corresponds to the rate reduction that yields $\mathbb{P}_{\mathrm{f}_1}<\zeta$. The simulation parameters are $\lambda=20$ packets/second, ${\underset{i}{\mathrm{min}} \{\mu_i-\lambda'_i\}}=25$ packets/second, $C=4.46$  nats/second (see~\eqref{eq:c}), $T=3600 \times2$ seconds, $\zeta=0.1$, $\epsilon=0.1$, $m=9$ (see~\eqref{eq:11}). The number of trials is $10,000$. As expected, larger values of $r'$ yields a smaller probability of failure for Alice. Although only the value of $m$ is tied to Scenario 1, this result applies to all scenarios with minor modifications.  

Note that the ideal reduced rate in the first phase ($\lambda -r' \epsilon \sqrt{2\lambda/(mT_1)} $ with $r'=1$) is expected to achieve $\mathbb{P}_{\mathrm{f}}\leq \zeta=0.1$. Although the simulation for $r'\in[0.6,1]$ has not been done due to processing time limits, the Fig.~\ref{fig:pf} shows that even with less rate reduction ($r'=0.5$) and hence less buffering, we achieve a much smaller probability of failure ($\mathbb{P}_{\mathrm{f}_1}\leq \zeta=2 \times 10^{-4}$). So, our buffering requirements are conservative rate reduction in the first phase is conservative. That leads to allocating a large portion of $T$ to the first phase, and a small portion to the second phase. The plot shows that in practice we can reduce the rate in the first phase less, and allocate a smaller portion of $T$ to the first phase.  

\subsection{Bob's decoding error probability} \label{sec:be}
We consider a single link and plot Bob's error probability ($\mathbb{P}_{\mathrm{f}_2}$), i.e., the probability that Bob extracts a wrong fingerprint from the link. Although the simulation results presented here are according to the number of links $m$ derived from Scenario 1, this result also applies to all scenarios with minor modifications. The simulation parameters are $\lambda=5.485$ packets/second, ${\underset{i}{\mathrm{min}} \{\mu_i-\lambda'_i\}}=5.5$ packets/second, $C=0.015$  nats/second (see~\eqref{eq:c}), $T=3600 \times 40$ seconds, $\zeta=0.15$, $\epsilon=0.1$, $m=13$ (see~\eqref{eq:11}). The number of trials is $2,000$. 

\begin{figure}
	\begin{center}
	\includegraphics[width=\textwidth,height=\textheight,keepaspectratio]{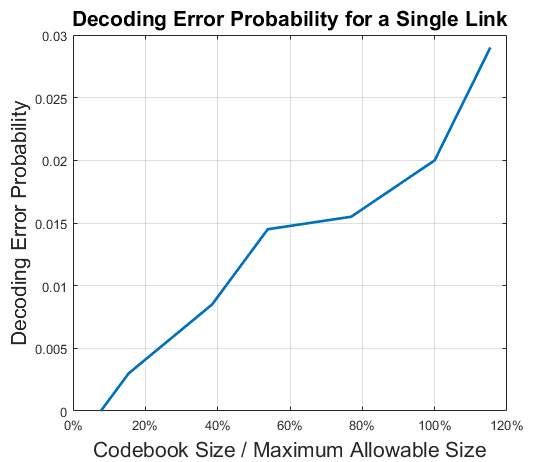}
	\end{center}
	\caption{The probability that Bob extracts a wrong fingerprint from a flow. Bob looks at the packet timings of the flow and extracts the fingerprint from it according to the codebook shared with Alice. The size of the codebook is $r'' \times m$, where $m$ is the ideal codebook size according to Theorem~\ref{thm:allflows}.}
	\label{fig:pe}
\end{figure}

The maximum allowable size of the codebook is $m=13$. For simulation, we let the size of the codebook be $\lfloor r'' \times m \rfloor$, where $r''\in [0.002,1.2]$ (see Fig.~\ref{fig:pe}). The x-axis is $r''$. According to Theorem~\ref{thm:allflows}, $r''=1$ corresponds to the ideal codebook size that results arbitrarily small error probability for Bob. Note that the results of Theorem~\ref{thm:allflows} is based on Shannon's random coding which relies on large $T$. If we consider larger values for $T$, we expect to see small error probabilities for Bob's decoding when the size of the codebook is ideal or smaller than that $r''\leq 1$. Currently, because of processing time limits, we observe $\mathbb{P}_{\mathrm{f}_2}=0.02$ when $r''=1$.

Although $T=3600 \times 40$ seconds, the length of the second phase is only $173$ seconds. In other words, if Alice and Bob are given $40$ hours, they only use about $3$ minutes of that time to embed and extract the fingerprints, and Alice uses the rest of the time to buffer packets in the first phase to ensure her fingerprinting will be successful. As stated in Section~\ref{sec:pa}, this is because the parameters for packet buffering are conservative. Improving the parameters and reducing the amount of time needed for buffering lies beyond the scope of this work since our primary goal is to establish the fundamental limits. Noting that using only $3$ minutes for embedding and extracting the fingerprints results in a decoding probability of error $\mathbb{P}_{\mathrm{f}_2}=0.02$, we can state that our current scheme is efficient in this way.      

\subsection{Robustness against processing time of queues}
Bob's detector relies on the fact that the queues are $M/M/1$ which implies the processing times of the queues are i.i.d exponential random variables. Here, we consider $M/G/1$ queues, whose processing times are i.i.d. samples of non-exponential random variables, and plot Bob's decoding error probability. We let the processing times of the queue be i.i.d. instantiations of a Weibull distribution with shape parameters $1$, $2$, $3$, $4$, with the same processing rate, $mu$. Note that the shape parameters $1$ corresponds to an exponential random variable.

\begin{figure}
 	\begin{center}
 		\includegraphics[width=\textwidth,height=\textheight,keepaspectratio]{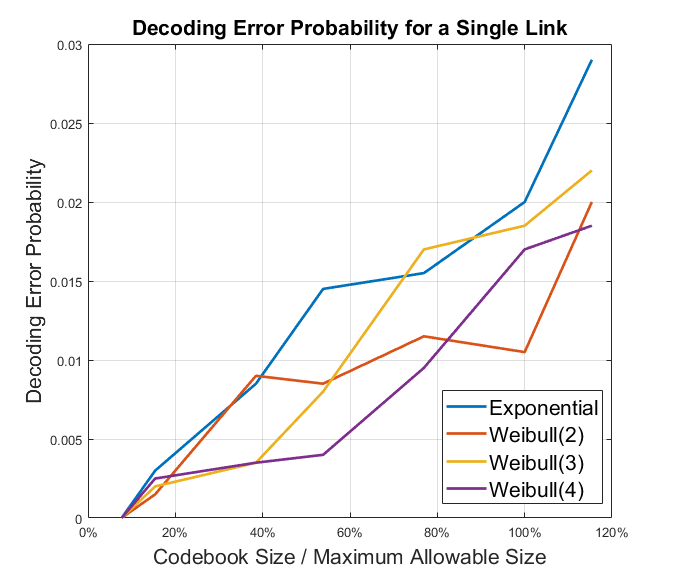}
 	\end{center}
 	\caption{The probability that Bob extracts a wrong fingerprint from a flow when the service times of the queue are i.i.d. instantiations of exponential distribution and Weibull distribution with shape parameters $2$, $3$, $4$ . Bob looks at the packet timings of the flow and extracts the fingerprint from it according to the codebook shared with Alice. The size of the codebook is $r'' \times m$, where $m$ is the ideal codebook size when the processing times are instantiations of an exponential random variable, according to Theorem~\ref{thm:allflows}.}
 	\label{fig:pe2}
 \end{figure}

Similar to Section~\ref{sec:be}, we consider a single link and plot Bob's error probability ($\mathbb{P}_{\mathrm{f}_2}$), i.e., the probability that Bob extracts a wrong fingerprint from the link (see Fig.~\ref{fig:pe2}). Although the simulation results presented here are according to the number of links $m$ derived from Scenario 1, this result also applies to all scenarios with minor modifications. The simulation parameters are the same as those of Section~\ref{sec:be}. According to Fig.~\ref{fig:pe2}, the change of distribution does not yield a major change in Bob's error probability, and thus Bob's decoder is robust against this change, i.d., if the distribution of the processing times of the queue changes from Weibull with shape parameter $1$ to Weibull with shape parameter $2,3,4$.

\section{Discussion} \label{disc}
\subsection{Source of the gain in Scenarios 2, 4, and 6}
Comparing the results of Scenarios~1, 3, and~5 (Alice fingerprints all flows she observes) with those of Scenarios~2, 4, and~6 (Alice fingerprints a subset of flows she observes), we notice a large gain for the number of flows that can be fingerprinted when Alice fingerprints the flows with a small probability. Intuitively, if $H_1$ is true, in Scenarios~1, 3, and~5, Willie is certain that there is only one possibility: all flows are slowed down by Alice in the first phase. However, if $H_1$ is true, the number of possible sets of flows that might have been slowed down by Alice in the first phase is ${m \choose M}$ for Scenarios~2 and~4, and $2^M$ for Scenario~6, where sets whose cardinality is about $Mq$ are more probable. Since a small portion of the flows is fingerprinted in Scenarios~2, 4, and~6, Willie needs to investigate a large number of flows to look for the decreasing of flow rates of a relatively (very) small random subset of those flows. This makes invisibility much easier to achieve and leads to the significant gains observed. 
\subsection {Alternative characterization of $m$ with respect to $M$ for Scenarios~2,4, and 6}
Consider Scenario~2 (Theorem~\ref{thm:pflows}). An alternative way to show the relation between the maximum number of flows $m$ that could be traced from a set of flows of size $M$ observed by Alice is:
\begin{itemize}
	\item If there exists a constant $\xi<C$ such that $M=\mathcal{O}(e^{ T \xi })$, then $m=\mathcal{O}(\sqrt{M})$.
	\item If for all $\xi<C$, $M=\omega(e^{ \xi T})$, then $m=\mathcal{O}(e^{TC_5}\sqrt{M})$, for all $C_5\in(0,C)$.
\end{itemize} 
This applies to Scenario~4 (Theorem~\ref{thm:pflowsd}) and Scenario~6 (Theorem~\ref{thm:pflowsw}), replacing $C$ with $C'$ and $C''$, respectively.

\subsection{Alice's knowledge about effective service times of the queues}
We presented results assuming Alice knows the effective service rates of the queues, i.e., $\mu_i-\lambda_i'$ for all $1\leq i\leq M$. We can show that if Alice does not know the effective service rates, but she knows a positive lower bound on each of them, then we achieve the same big-O results for the number of flows that Alice and Bob can trace. Furthermore, if she does not know the lower bounds, our big-O results achieved for Scenarios~1,~3, and~5 will change to Little-o results. 
\subsection{Sharing the fingerprinting codebook}
The use of a secret pre-shared key has been largely addressed in security and cryptography~\cite{katz1996handbook,stinson2005cryptography}. In practice, the distribution of secret keys can be done by face-to-face meeting, use of a trusted courier, or sending the key through an existing encryption channel. In many scenarios a secure low throughput channel is available that the parties can use to share the key. Also, Diffie-Hellman key exchange (DH) can be used for sharing such a key~\cite{steiner1996diffie} over a public channel.

\subsection{Delay performance}
Our fingerprinting scheme requires that Alice first buffers packets, which increases the end-to-end delay of the network. We can show that the average packet delay in Scenarios 1, 3, and 5 is $\mathcal{O}(\sqrt{\log T})$, and in Scenarios 2, 4, and 6 is $\mathcal{O}(\sqrt{T})$.

We have shown in the reliability analyses that the packet delay does not impact Bob's decoding, and he can extract fingerprints with arbitrarily small error probability. This is true because Bob extracts the fingerprints from inter-packet delays. Furthermore, it does not help Willie's detection. In other words, in the invisibility analysis we have shown that although packets experience delays, Willie cannot detect Alice and Bob's fingerprinting. This is true because Willie does not have access to the original packet timings; rather, he only knows the statistics of the packet timings which change only slightly and are undetectable to him. Consider the users of the network. Although this delay is not tolerable in applications such as voice over IP, there are many applications such as file transfer that allow for this.

\subsection{Unwinding packets in Alice's buffer}
Note that Alice's fingerprinting requires that she first buffers packets. We can show that Alice will have $\mathcal{O}(\sqrt{T})$ packets in her buffer after the second phase ends at $t=T$. To unwind the packets, after $t=T$, Alice relays all the flows she receives at the rate she receives them, and insert packets from her buffer according to a Poisson process of rate $\Delta$. Similar to the arguments where we showed that the change of rate from $\lambda-\Delta$ is undetectable to Willie, we show that the change of rate from $\lambda$ to $\lambda+\Delta$ is undetectable to Willie, and thus Alice's unwinding is invisible. Similar analyses has been addressed in our previous works~\cite{soltani2015covert,soltani2016allerton}. 

\section{Future work} \label{sec:fw}
The future work consists of alternative network models and extending the current network model. We will consider the cases where 1) packets drop; 2) packets are duplicated; 3) the order of the packets change; 4) packets are fragmented; and 5) flows are re-packetized. In addition, we will apply the results of~\cite{soltani2016allerton} to extend our results to $G/M/1$ queues and we will consider other queuing models. Furthermore, we will apply the work of~\cite{mimcilovic2006mismatch} to consider a network of parallel links where each link contains a set of $M/M/1$ single input/output queues in tandem, and then we will extend this to tandem queues shared between a flow between Alice and Bob (main flow) and independent interfering flows. Furthermore, we will use~\cite[Corollary 3.3]{kelly2011reversibility} to relax the condition of independent interference for queues on each route. Moreover, we will extend our model to a feedforward multiclass product form network~\cite{baskett1975open} containing parallel links where each link consists of multiple $M/M/1$ queues in tandem shared between a flow between Alice and Bob (main flow) as well as interfering flows.

\section{Conclusion} \label{sec:con}
We have presented the construction and analysis for invisible fingerprinting of flows to infer the connections between input and output links of a network that is modeled as $M$ independent, parallel, and work-conserving $M/M/1$ queues with background traffic. In a setting where flows whose packet timings are governed by Poisson processes visit Alice, Willie, the network, and Bob respectively, we have presented a construction where Alice fingerprints flows in a time interval of length $T$ by manipulating packet timing of the flows according to a fingerprint codebook shared with Bob and unknown to Willie. In particular, each codeword (fingerprint) of the codebook is a unique flow identifier which corresponds to a sequence of inter-packet delays. If flow rates are equal, Bob observes only flows with fingerprints, and Alice chooses to fingerprint all $M$ input flows of the network that she observes, Alice and Bob can invisibly trace the fingerprinted flows as long as $m=M=\mathcal{O}(T/\log T)$. But, if she fingerprints a subset of the flows $\mathcal{F}_{f}$, Alice and Bob can invisibly trace the fingerprinted flows as long as $m=|\mathcal{F}_{f}|=o(\min\{\sqrt{M},e^{TC_1}\})$, for all $C_1\in(0,C)$, with more accurate characterizations of $m$ with respect to $M$ presented in~\eqref{eq:bigM} and~\eqref{eq:1}. Similar results hold for arbitrary flow rates as well as the case where Bob observes flows with and without fingerprints, with minor modifications.

\makeatletter
\renewcommand{\theparagraph}{\Alph{paragraph}}
\renewcommand{\@seccntformat}[1]{\csname the#1\endcsname.\quad}
\makeatother

\appendix

\paragraph{\textbf{Applicability of covertness metric when $\mathbb{P}(H_0)\neq\mathbb{P}(H_1)$}}\label{ap.1}

Definition~\ref{def:inv} implies that when $\mathbb{P}(H_0)=\mathbb{P}(H_1)=1/2$, Alice can make Willie's detector operate as close as desired to a detector that disregards Willie's observation, e.g., tosses a fair coin to decide whether Alice is fingerprinting. 

For $\mathbb{P}(H_0)\neq\mathbb{P}(H_1)$, if Alice's scheme satisfies the invisibility metric in Definition~\ref{def:inv}, she can also make Willie's detector operate as close as desired to a detector that disregards Willie's observations, as follow. Recall that $\mathbb{P}_{\mathrm e}^{(\mathrm w)} = \frac{\mathbb{P}_{\mathrm {FA}}+\mathbb{P}_{\mathrm {MD}}}{2}$ is Willie's error probability when prior probabilities are equal, $\mathbb{P}(H_0) = \mathbb{P}(H_1)=0.5$. Denote by $\mathbb{P}_{\mathrm e}'^{(\mathrm w)}$ Willie's error probability when prior probabilities are not equal. Then:
\begin{align}
\nonumber \mathbb{P}_{\mathrm e}'^{(\mathrm w)}&=(1-\mathbb{P}({H_1}))\mathbb{P}_{\mathrm {FA}}+\mathbb{P}({H_1})\mathbb{P}_{\mathrm {MD}},\\
\nonumber &\geq 2 \min{(\mathbb{P}({H_1}), 1-\mathbb{P}({H_1}))}\frac{\mathbb{P}_{\mathrm {FA}}+\mathbb{P}_{\mathrm {MD}}}{2},\\
\label{eq:11111} &\geq 2 \min{(\mathbb{P}({H_1}), 1-\mathbb{P}({H_1}))}\mathbb{P}_{\mathrm e}^{(\mathrm w)},
\end{align}
By Definition~\ref{def:inv}, if Alice's fingerprinting is invisible, then for large enough $T$ she can achieve $\mathbb{P}_{\mathrm e}^{(\mathrm w)}>\frac{1}{2}-\epsilon$, for all $\epsilon>0$. Hence,~\eqref{eq:11111} yields:
\begin{align}
\nonumber \mathbb{P}_{\mathrm e}'^{(\mathrm w)}&\geq  \min{(\mathbb{P}({H_1}), 1-\mathbb{P}({H_1}))} (1-2 \epsilon),\\
\label{eq:init}  &\geq \min{(\mathbb{P}({H_1}), 1-\mathbb{P}({H_1}))} - \epsilon',
\end{align}
\noindent where $\epsilon'=2 \epsilon \min{(\mathbb{P}({H_1}), 1-\mathbb{P}({H_1}))}$. Consider a detector that disregards Willie's observations: if $\mathbb{P}(H_0)>0.5$, Willie always decides that Alice is fingerprinting; otherwise, Willie decides that she is not. Using this detector, Willie achieves $\mathbb{P}_{\mathrm e}'^{(\mathrm w)}=\min(\mathbb{P}(H_1),1-\mathbb{P}(H_1))$. From~\eqref{eq:init}, Alice can make Willie's detector operate as close as desired to this detector.

\paragraph{\textbf{Proof of~\eqref{eq:3}}}\label{ap.2}
Denote by $\mathbb{P}_0$ the pdf for Willie's observations in the first phase under the null hypothesis $H_0$ (Alice is not fingerprinting), and by $\mathbb{P}_1$ the joint pdf for corresponding observations under the hypothesis $H_1$ (Alice is fingerprinting) in the first phase. Note that under $H_1$, Alice in the first phase slows down the flow $f_i$ from rate $\lambda_i$ to $\lambda_i-\Delta_i$, for $1\leq i\leq m$. Since the number of observed packets for Poisson processes is a sufficient statistic for hypothesis testing~\cite{soltani2015covert}, 
\begin{align}
\nonumber \mathbb{P}_0&= \prod_{i=1}^{m} \mathbb{P}_{\lambda_i}(n_i),\\
\nonumber \mathbb{P}_1&=\prod_{i=1}^{m} \mathbb{P}_{\lambda_i-\Delta_i}(n_i),
\end{align}
where $\mathbb{P}_{\lambda}(n)$ is the probability mass function (pmf) of the number of packets in time $T_1$ for a flow whose packet timings are governed by a Poisson process with rate $\lambda$, and $T_1$ is the length of the first phase. 
Observe
\begin{align}
\nonumber  &\mathcal{D}(\mathbb{P}_{\lambda_i-\Delta_i}(n_i)||\mathbb{P}_{\lambda_i}(n_i)) \\ \label{eq:02} &=\Delta_i T_1 -(\lambda_i -\Delta_i) T_1 \log {\frac {\lambda_i }{\lambda _i-\Delta_i}} \leq  \frac{T_1 \Delta_i^2}{2(\lambda_i-\Delta_i)},
\end{align}
\noindent where the last steps follows from the inequality $\ln(1+x)\geq x-x^2/2$ for $x\geq 0$. Thus, 
\begin{align}
\nonumber \mathcal{D}(\mathbb{P}_1||\mathbb{P}_0)= \sum_{i=1}^{m} \mathcal{D}(\mathbb{P}_{\lambda_i-\Delta_i}(n_i)||\mathbb{P}_{\lambda_i}(n_i))\leq  \sum_{i=1}^{m} \frac{T_1 \Delta_i^2}{2(\lambda_i-\Delta_i)}.
\end{align}
\noindent Let $\Delta_i = \epsilon \sqrt{\frac{2 \lambda_i}{m T_1}}$, where $\epsilon>0$. Therefore, $\mathcal{D}(\mathbb{P}_1||\mathbb{P}_0)\leq  \frac{\epsilon^2 }{m} \sum_{i=1}^{m} \frac{\lambda_i}{\lambda_i- \sqrt{2 \lambda_i/T_1}}$. For large enough $T_1$, $\frac{\lambda_i}{\lambda_i- \sqrt{2 \lambda_i/T_1}}\leq 2$, and thus $\mathcal{D}(\mathbb{P}_1||\mathbb{P}_0) \leq  2 \epsilon^2$ as $T_1 \to \infty$. Combining with~\eqref{eq:0}, $\mathbb{P}_{\mathrm e}^{(\mathrm w)} \geq \frac{1}{2}-\frac{\epsilon}{2}\geq\frac{1}{2}-{\epsilon}$. Consequently, the first phase is invisible.

\paragraph{\textbf{Proof of~\eqref{eq:29}}} \label{ap.0} Consider the following fact:
\begin{fact} \label{f:1} For $x,y>0$, if $x<y/W(y)$,  then $x \log x < y$.  
\end{fact}
\noindent \textbf{Proof.} Assume $x'=y/W(y)$. First, we show that $x' \log x' = y$. From the definition of the Lambert-W function, $W(y) e^{W(y)} = y$. Therefore, $ W(y) = \log \frac{y}{W(y)}$. Consequently, 
\begin{align}
\label{eq:123} x' \log x'= \frac{y}{W(y)} \log{\frac{y}{W(y)}} = \frac{y}{W(y)} W(y) = y. 
\end{align}
\noindent Since $x'=y/W(y)$, $x<y/W(y)$ implies that $x<x'$. Because $x \log x$ is an increasing function of $x$, $x \log x< x' \log x' = y$, and the proof is complete.
\QEDB 

Next, for both cases $m\geq 1+m \alpha$ and $m< 1+m \alpha$ we show that $TC>(1+m \alpha) \log m$, which implies~\eqref{eq:29}. Consider $m\geq 1+m \alpha$. Note that~\eqref{eq:11} implies $m < \frac{TC}{W(TC)}$. Therefore, Fact~\ref{f:1} yields:
$$ TC> m \log{m}.$$ 
Since $m\geq 1+m \alpha$,
$$ TC> m \log{m} >(1+m \alpha) \log{m}.$$ 

\noindent Now, consider $m< 1+m \alpha$. Note that~\eqref{eq:11} implies that $m< {{\alpha}^{-1}}\left({\frac{TC}{W(TC)}}-1\right)$, which implies  $1+ m \alpha<\frac{TC}{W(TC)}$ Hence, ~Fact~\ref{f:1} yields 
\begin{align}
\nonumber TC &> (1+m \alpha) \log{(1+m \alpha)}\geq (1+m \alpha) \log{m},
\end{align}
\noindent where the last inequality follows from $m< 1+m \alpha$. Consequently,~\eqref{eq:11} satisfies~\eqref{eq:29}.

\paragraph{\textbf{Proof of~\eqref{eq:1234}}}\label{ap.4} Observe:
\begin{align}
\nonumber \mathcal{D}(\mathbb{P}_1||\mathbb{P}_0) &\stackrel{(a)}{=} M \mathcal{D}\left(p\mathbb{P}_{\lambda-\Delta}(n)+(1-p)\mathbb{P}_{\lambda}(n)||\mathbb{P}_{\lambda}(n)\right),\\
\nonumber &\stackrel{(b)}{=} M \mathbb{E}_{1} \left[\ln\left(\frac{p\mathbb{P}_{\lambda-\Delta}(n)+(1-p)\mathbb{P}_{\lambda}(n)}{\mathbb{P}_{\lambda}(n)}\right)\right],\\
\nonumber &= M \mathbb{E}_1 \left[\ln\left(p e^{\Delta T_1} \left(\frac{ \lambda -\Delta  }{\lambda  }\right)^n+(1-p)\right)\right],\\
\nonumber & \stackrel{(c)}{\leq}   M p\mathbb{E}_1 \left[ e^{\Delta T_1} \left(\frac{ \lambda -\Delta  }{\lambda  }\right)^n\right]-Mp,\\
\label{eq:13} & \stackrel{(d)}{=} M p^2    ( e^{ \Delta^2 T_1/\lambda}-1)\stackrel{(e)}{=} \epsilon^2/2.
\end{align}
where $(a)$ follows from the chain rule for relative entropy~\cite[Eq. (2.67)]{cover2012elements}, $\mathbb{E}_1[\cdot]$ denotes expected value with respect to the pdf $\left(p\mathbb{P}_{\lambda-\Delta}(n_i)+(1-p)\mathbb{P}_{\lambda}(n_i)\right)$, $(b)$ follows from the definition of the Kullback–Leibler divergence, $(c)$ is true since $\ln(1+x)\leq x$, $(d)$ is true since 
\begin{align}
\nonumber \mathbb{E}_1 \left[  \left(\frac{ \lambda -\Delta  }{\lambda  }\right)^n\right]
=e^{- \Delta T_1}\left(p  e^{ \Delta^2 T_1/\lambda} + (1-p) \right),
\end{align}
\noindent and $(e)$ follows from substituting the values of $\Delta$, $p$, and $T_1$ given in~\eqref{eq:del2},~\eqref{eq:p1},  and~\eqref{eq:lph1} respectively.

\paragraph{\textbf{Proof of~\eqref{eq:4}}}\label{ap.5} The first difference is in the number of packets that Alice can buffer from each flow in the first phase. Here, since Alice slows down each flow from rate 
$\lambda$ to $\lambda-\Delta$, where $\Delta$ is given in~\eqref{eq:del2}, the probability that Alice can buffer more than $\Delta T_1/2$ packets in the second phase tends to one as $T \to \infty$. Therefore, letting $t=T_1$  and $k=\Delta T_1/2 =  \sqrt{{\lambda}{T_1} \ln\left(1+\frac{\epsilon^2 M }{2 m^2}\right)}/4$ in~\eqref{eq:125} yields:
\begin{align}
\label{eq:134}\lim\limits_{T \to \infty}  P_{\mathrm{f}_1}\leq 1-\lim\limits_{T \to \infty} \mathrm{erf}\left(\sqrt{\frac{ {T_1} \ln\left(1+\frac{\epsilon^2 M}{2 m^2}\right)}{8  T_2}}\right).
\end{align}
\noindent The second difference in the analysis of $\mathbb{P}_{\mathrm{f}_1}$ is due to differences in the expressions for $T_1$ and $T_2$. By~\eqref{eq:lph1} and~\eqref{eq:lph2}, $T_1/T_2 =\alpha'/\ln(1+\frac{\epsilon^2 M}{2 m^2})$. Therefore,~\eqref{eq:134} yields:
\begin{align}
\label{eq:11110} \lim\limits_{T \to \infty} P_{\mathrm{f}_1}\leq 1-\mathrm{erf}\left(\sqrt{\frac{ {{\alpha'} } }{8}}\right)= 1-\mathrm{erf}\left(\epsilon\sqrt{\frac{ {{\alpha} } }{8}}\right), 
\end{align}
\noindent where the last step is true since $\alpha'=\epsilon^2 \alpha$. By~\eqref{eq:alpha},

\paragraph{\textbf{Proof of~\eqref{eq:1}}}\label{ap.3}
If $M=\mathcal{O}(1)$, by~\eqref{eq:bigM}, the left hand side (LHS) of~\eqref{eq:1} is $m=M=\mathcal{O}(1)$. Now, consider the right hand side (RHS) of~\eqref{eq:1}. Since $m=O(1)$, there exists $\rho$ such that for large enough $T$, $m\leq \rho$. Consequently, the RHS of~\eqref{eq:1} is $\Omega(e^{\frac{TC}{1+\rho'}})$, where $\rho'=\alpha'/\ln(1+{\frac{\epsilon^2  }{2 \rho}})$. Thus,~\eqref{eq:1} is satisfied.

If $M=\omega(1)$ and $M=\mathcal{O}(e^{2 TC})$, $m = o(\min{\sqrt{M},e^{TC_1}})$, where $C_1\in(0,C)$. Thus, the LHS of~\eqref{eq:1} is $o(e^{T C_1})$. To show~\eqref{eq:1} is satisfied, it suffices to show that there exists a constant $C_1'\in (C_1,C)$ such that makes the RHS of~\eqref{eq:1} $\Omega(e^{TC_1'})$, which is true since
\begin{align}
\label{eq:2}\exp\left( \frac{TC }{1+\alpha'/\ln(1+{\frac{\epsilon^2 m }{2 M^2}})}\right) \geq e^{ {TC }{\left(1-\alpha'/\ln(1+{\frac{\epsilon^2 M }{2 m^2}})\right)}} 
\end{align}
\noindent provided that $1/(1+x)\geq 1-x$ for all $x>0$. Note that $m = o(\min{\sqrt{M},e^{TC_1}})$ implies that $m\in o(\sqrt{M})$, and thus $\alpha'/\ln(1+{\frac{\epsilon^2 M }{2 m^2}})$ in the RHS of~\eqref{eq:2} gets as small as desired.

If $M=\omega(e^{2 TC})$, $m=\Theta(e^{TC_2})$ for any $C_2 \in (0,C)$, and thus the LHS of~\eqref{eq:1} is $\Theta(e^{TC_2})$. Now, consider the RHS of~\eqref{eq:1}. Since $m=\Theta(e^{TC_2})$ and $M=\omega(e^{2 TC})$, $m=o(\sqrt{M})$, and thus $\alpha'/\ln(1+{\frac{\epsilon^2 M }{2 m^2}})$ in the RHS of~\eqref{eq:2} gets as small as desired. Consequently, there exists a constant $C_2'\in (C_2,C)$ such that the RHS of~\eqref{eq:1} is $\Omega(e^{TC_2'})$. Hence, ~\eqref{eq:1} is satisfied. 
\newpage
\bibliographystyle{ieeetr}

\end{document}